\documentclass[superscriptaddress,aps,amsmath,amssymb,showpacs,showkeys]{revtex4}
\usepackage[dvips]{graphicx,color}
\usepackage{times}
\usepackage{xcolor}
\usepackage[%
  colorlinks=true,
  urlcolor=blue,
  linkcolor=red,
  citecolor=blue
]{hyperref}
\usepackage{orcidlink}
\usepackage{mathrsfs}
\usepackage{amsmath}
\usepackage{accents}
\usepackage{mathtools}
\usepackage{amssymb}
\usepackage{graphicx}
\usepackage{epsfig}
\usepackage{dcolumn}
\usepackage{bm}
\begin{document}
\title{Investigating the effects of gravitational lensing by Hu-Sawicki $\boldsymbol{f(R)}$ gravity black holes}

\author{Gayatri Mohan \orcidlink{0009-0008-0654-5227}}
\email[Email: ]{gayatrimohan704@gmail.com}

\affiliation{Department of Physics, Dibrugarh University,
Dibrugarh 786004, Assam, India}

\author{Nashiba Parbin \orcidlink{0000-0002-3570-0117}}
\email[Email: ]{nashibaparbin91@gmail.com}

\affiliation{Department of Physics, AMC Engineering College,
Bangalore 560083, Karnataka, India}
\affiliation{Visvesvaraya Technological University (VTU), Belagavi 590018, Karnataka, India}

\author{Umananda Dev Goswami \orcidlink{}}
\email[Email: ]{umananda@dibru.ac.in}

\affiliation{Department of Physics, Dibrugarh University,
Dibrugarh 786004, Assam, India}

\begin{abstract}
In this work, gravitational lensing in the weak and strong field limits is 
investigated for black hole spacetime within the framework of Hu-Sawicki 
$f(R)$ gravity. We employ the Ishihara {\it et al.}~approach for weak lensing 
and adopt Bozza’s method for strong lensing to explore the impact of 
Hu-Sawicki model parameters on lensing phenomenon. The deflection angles are 
computed and analyzed in both the field limits. Our investigation in the weak 
as well as the strong lensing reveals that in the case of Hu-Sawicki black 
holes, photons exhibit divergence at smaller impact parameters for different 
values of the model parameters compared to the Schwarzschild scenario and the 
photon experiences negative deflection angle when impact parameter moves 
towards the larger impact parameter values. Additionally, by calculating 
strong lensing coefficients we study their behavior with model parameters. 
The strong lensing key observables associated with the lensing effect viz.~the 
angular position $\vartheta_{\infty}$, angular separation $s$ and relative 
magnification $r_\text{mag}$ are estimated numerically by extending the 
analysis to supermassive black holes $\text{SgrA}^*$ and $\text{M87}^*$ 
and analyzed their behavior concerning the parameters for each black hole. 
The analysis shows that $\text{SgrA}^*$ demonstrates larger values of 
$\vartheta_{\infty}$ and $s$ relative to $\text{M87}^*$.
\end{abstract}
	
	
\maketitle
\section{Introduction}
\label{sec.1}
The predictions of black holes (BHs) and gravitational waves (GWs) are the two 
most incredible and noteworthy achievements of Einstein's theory of General 
Relativity (GR), published in $1915$ \cite{Einstein_1915}. 
The historic triumph of the Event Horizon Telescope (EHT) in capturing the 
first photographs of the supermassive BHs M$87^*$ \cite{AkiyamaM87_2019}
and SgrA$^*$ \cite{AkiyamaSgrA_2022} have unfurled new aspects for studying BHs 
thereby fetching more attention as one of the active areas in modern 
astrophysics. In essence, the study of the optical features of BHs has prompted 
considerable interest amongst the scientific population. This dynamic is 
persistently advancing with enhanced technologies and the accumulation of 
fresh data. In passing it needs to be mentioned that the detection of GWs by 
the LIGO detector system in 2015 \cite{Abbott_2016,Abbott_2017} is another significant milestone of GR 
on its avenue of success, which unfolded different ways of looking into the 
Universe.   

EHT's BH images and detection of GWs by LIGO have successfully corroborated 
predictions of Einstein's GR, simultaneously raising queries on the 
consequential challenges faced by it, which include deciphering the expansion 
history of the Universe \cite{Reiss_1998, Perlmutter_1999}, the observed 
rotational dynamics of the galaxies \cite{Rubin_1970, Young_2017, 
Nashiba_2023a,2024_mohan,2024_gayatri}, the need for exotic stuff such as dark matter and dark energy 
\cite{Bertone_2018, Swart_2017, Garret_2011}, existence of spacetime 
singularities within BHs \cite{Hawking_1967}, the large scale structure 
\cite{Hawking_1973}, etc. These led to an extensive investigation of different 
theories of gravity beyond the GR framework, known as modified theories of 
gravity (MTGs) \cite{Clifton_2012, Clifton_2006}, where higher curvature terms 
or extra fields are incorporated in the Einstein-Hilbert (EH) action. In recent 
times, a plethora of MTGs \cite{Capozziello_2011, Nojiri_2011, Nojiri_2011a, 
Nojiri_2017, Faraoni_2004, Moffat_2006, Bahamonde_2015, Bahamonde_2023, 
Canfora_2021} have been introduced which impart richer frameworks to comprehend 
gravity in a better way. $f(R)$ theory of gravity \cite{Sotiriu_2010, 
Felice_2010} is one of the most straightforward extensions that lays out a 
constructive way to interpret the basic principles and limitations behind the 
modification of GR. It has been 
under detailed investigation over the years \cite{Olmo_2011, Vikram_2018, 
Sharma_2020, Nashiba_2021, Naik_2018, Shirasaki_2018, Errehymy_2024, 
Guilabert_2024, Talukdar_2024} and is categorized under those theories that 
encompass higher-order curvature invariants. Currently, a few models of $f(R)$ 
gravity have been proposed. The Starobinsky model \cite{Starobinsky_2007}, the 
Hu-Sawicki model \cite{Sawicki_2007}, the Tsujikawa model \cite{Cen_2019}, and 
another recently introduced model \cite{Dhruba_2020} are a few of the known 
viable $f(R)$ gravity models. In literature, these models have been rigorously 
studied in various aspects such as unraveling the mystery of dark matter 
\cite{Nashiba_2021, Katsuragawa_2018}, puzzles of the early Universe 
\cite{Katsuragawa_2019}, cosmological and astrophysical consequences of these 
models \cite{Bessa_2022, Jyatsnasree_2022}, and so on.

In $1919$, Arthur Eddington, Frank Dyson and Charles Davidson detected the 
deflection in the path of light emerging from stars in the Hyades cluster, 
around the sun during a solar eclipse \cite{Dyson_1920}. This groundbreaking 
observation served as the first experimental proof of Einstein's GR and thus 
emerged the beautiful phenomenon of gravitational lensing 
\cite{Schneider_1992, Schneider_2006}. Extensively studied in cosmology and 
astronomy \cite{Bartelmann_2010, Wambsganss_1998, Blandford_1992, Congdon_2018},
lensing occurs as light from a distant source gets deviated in the vicinity of 
massive objects, such as BHs, galaxy clusters, etc. The deflection 
angle allows us to study the optical properties of such massive objects. BHs 
exhibit exceptional laboratories to explore strong gravitational effects. 
Implementing the magnificent phenomenon of gravitational lensing on BHs, it is 
feasible to classify a variety of BH models 
\cite{Eiroa_2002, Gyulchev_2007, Zhao_2016, Zhao_2017, Kumar_2020, Dong_2024, 
Nashiba_2023b} and also to test different theories of gravity 
\cite{Nashiba_2023b, Nashiba_2023c, Nashiba_2024, Sahu_2015, Soares_2023, 
Badia_2017, Vagnozzi_2022, Pantig_2023}. Gravitational lensing can also act as 
a robust astrophysical tool for the study of the gravitational field of massive 
objects as well as for uncovering the mystery of dark matter 
\cite{Clowe_2006, Massey_2010, Fu_2022}. 

When numerous photons reach the vicinity of BHs, the effects of 
gravitational lensing can be observed which contribute towards the formation of 
distinct features of the BH such as its shadow, relativistic images at 
the event horizon, and a photon ring. To figure out these gravitational field 
features of the BH from the observational as well as theoretical 
standpoint, a pivotal role is portrayed by the study of the null geodesics 
around the BH. In the strong field limit, the foundational work 
contributed by Darwin \cite{Darwin_1959} to interpret the lensing effects 
marked the dawn of research in this area. Such studies in the strong field 
limit have gained considerable recognition in recent years as more information 
on BHs can be extracted from the same \cite{2002_Bozza, Bozza_2003, 
Whisker_2005, Chen_2009, Liu_2010, Ding_2011, Sotani_2015, Tsukamoto_2022, 
Tsukamoto_2023, Choudhuri_2023, Soares_2023a, Mustafa_2024, Kuang_2022}. 
Virbhadra and Ellis explored the Schwarzschild BH to investigate 
various aspects of gravitational lensing. They first formulated the lens 
equation in the strong field domain \cite{Virbhadra_2000}. Next, they analyzed 
the behavior of relativistic images for a Schwarzschild BH 
\cite{Virbhadra_2009, Virbhadra_2022, Virbhadra_2024}. They further explored 
the lensing by naked singularities \cite{Virbhadra_2002}. Many studies also 
involved the analysis of time delays caused due to gravitational lensing in the 
strong field limit \cite{Bozza_2004, Hsieh_2021, Islam_2021, Cavalcanti_2016, 
Lu_2016, Eiroa_2013, Zhu_2019}. Studies concerning the weak field limit come 
into play when we consider the photons to be at a significant distance from the 
BH. At this point, the strong field limit does not remain relevant. 
In $2008$, Gibbons and Werner ventured on the path to study the gravitational 
lensing features in the weak field limit \cite{Gibbons_2008} and proposed an 
alternative approach. They implemented the Gauss-Bonnet Theorem (GBT) 
\cite{Carmo_2016, Klingenberg_1978} to derive the deflection of light in the 
weak gravitational field of a static spherically symmetric spacetime. A number 
of articles have revealed that this approach can be utilized to deduce the 
deflection angle for various BH spacetimes \cite{Jususfi_2017, 
Sakalli_2017, Ovgun_2019, Gyulchev_2019, Kumar_2020a, Pahlavon_2024, Li_2020, 
Jha_2023, Panah_2020, Qiao_2023}. Subsequently, Werner developed their analysis 
of axially symmetric spacetimes implementing Finsler geometry \cite{Werner_2012,
 Jusufi_2018}, although this approach is found to be a bit challenging. A few 
years later, in $2016$, Ishihara and his co-workers developed the 
Gibbons-Werner approach further while considering finite distances 
\cite{Ishihara_2016}. Refs.~\cite{Ono_2017, Kumar_2019, Kumar_2020a, Zhu_2019, 
Crisnejo_2019} report the use of this extension for stationary spacetimes. 
Furthermore, for non-asymptotically flat BHs, the use of this extended 
approach is reported in Refs.~\cite{Ono_2019, Panpanich_2019, Takizawa_2020, 
Carvalho_2021, Nashiba_2023b, Nashiba_2024}. In the last decade, studies of 
gravitational lensing around naked singularities \cite{Atamurotov_2022, 
Chen_2024, Hossain_2024}, wormholes \cite{Shaikh_2019, Saurav_2024, 
Godani_2023} and other exotic objects \cite{Surajit_2023, Davies_2020} have 
also increased along with the study of lensing in spacetimes surrounded by 
dark matter \cite{Pantig_2022, Javed_2022}.

Motivated by the factors presented above, our research tends to investigate the 
lensing features of a BH within the framework of Hu-Sawicki $f(R)$
gravity model \cite{2024_karm}. In this theory, the BH spacetime has 
been recently deduced \cite{2024_karm} which makes it more intriguing to 
explore the 
gravitational bending elements in the weak as well as strong field limits. For 
the weak field lensing, the Ishihara \textit{et al.} approach is employed to 
analyze the effects of the Hu-Sawicki model parameters on the bending angle of 
light as it reaches the vicinity of such a BH. Next, the strong field 
lensing is studied by implementing the methodologies forwarded by Bozza 
\cite{2002_Bozza} and the effect of the model parameter on the lensing 
observables is also investigated. 

The remaining paper is organized in the following pattern. In Section
\ref{sec.2}, we give a brief explanation of the framework that is to be used 
for the proposed research. In Section~\ref{sec.3}, we deduce the deflection 
angle for the weak field limit. In Section~\ref{sec.4}, the gravitational 
bending angle of light is derived in the strong field limit of the BH 
spacetime. In addition, in this section, the lensing observables are also 
computed and analyzed in the $f(R)$ gravity framework. In Section~\ref{sec.5}, 
we summarize and conclude the findings of our research.

\section{Black holes in Hu-Sawicki $f(R)$ gravity theory}
\label{sec.2}

Modifying Einstein's GR is an intricate and challenging piece of work. As 
mentioned earlier, $f(R)$ gravity theory is one of the simplest modifications 
of GR and hence, is considered frequently for various astrophysical and 
cosmological studies. One compelling advantage of this theory is that it can 
avoid the Ostr\"{o}gradsky instability \cite{2021_Donoghue}. Indeed, 
Ostr\"{o}gradsky instability is a fundamental issue that arises in field 
theories when equations of motion involve higher-order derivatives. This 
instability gives rise to a ghost degree of freedom and results in a system 
with unbounded Hamiltonians, i.e.~the energy can become arbitrarily negative, 
causing runaway solutions and making the theory physically inconsistent. The 
fundamental cause of this instability lies in the Hamiltonian formulation of 
higher-derivative theories. It is a consequence of Ostr\"{o}gradsky’s theorem, 
which states that a non-degenerate Lagrangian containing higher than 
second-order time derivatives leads to a Hamiltonian that is not bounded 
from below \cite{2013_taijun}. 

It is well known that in order to deduce the field equations of $f(R)$ 
gravity, the first step is to rewrite the EH action by replacing the Ricci 
scalar $R$ with a function of $R$, usually designated as $f(R)$. The generic 
action that defines the $f(R)$ gravity theory is specified as 
\cite{Sotiriu_2010}
\begin{equation}
S = \frac{1}{2\kappa} \int d^4x \sqrt{-g} f(R) + S_m,
\label{eq1}
\end{equation}
where $\kappa = 8\pi G c^{-4}$ and the action of matter is denoted by 
$S_m$. Varying the above action with respect to the metric $g_{\mu\nu}$, we 
arrive at the field equations for $f(R)$ gravity as given by 
\begin{equation}
	FR_{\mu\nu} - \frac{1}{2}f(R)g_{\mu\nu} - (\nabla_\mu \nabla_\nu - g_{\mu\nu}\square)F = \kappa T_{\mu\nu},
	\label{eq2}
\end{equation}
where $F$ is obtained by differentiating $f(R)$ with respect to $R$, and 
$\square = \nabla_\alpha \nabla^\alpha$. The energy momentum tensor 
$T_{\mu\nu}$ is given as
\begin{equation}
	T_{\mu\nu} = \frac{-2}{\sqrt{-g}}\frac{\delta(\sqrt{-g}S_m)}{\delta g_{\mu\nu}}.
	\label{eq3}
\end{equation}
Trace of Eq.~\eqref{eq2} can be obtained as
\begin{equation}
	f(R) = \frac{1}{2} \big[3\,\square F + FR - \kappa T\big].
	\label{eq4}
\end{equation}
This equation contains fourth-order derivatives of the metric through 
$\square F$. It is a dynamical equation that suggests $F$ as an additional 
propagating degree of freedom in the theory. This extra degree of freedom 
accounts for the deviations from GR. 

In Ref.~\cite{2024_karm}, the authors have deduced a BH spacetime in the 
Hu-Sawicki model \cite{Sawicki_2007} of $f(R)$ gravity. This model represents 
one of the viable functional forms of the $f(R)$ gravity theory which is found 
to be consistent observationally in cosmological scales \cite{Martinelli_2009, 
Hough_2020}. In $2007$, Wayne Hu and Ignacy Sawicki came up with this model to 
interpret the present accelerating Universe excluding the cosmological 
constant. One of its interesting traits is that it is able to meet the 
requirements of the solar system tests thereby being valid in the local scales 
too. Presently, numerous research works are employing the Hu-Sawicki model to 
investigate diverse aspects of cosmology and astrophysics 
\cite{Bayron_2023, Kou_2024, Fakhry_2024, Lyall_2023}. The model is 
presented as 
\begin{equation}
	f(R) = -\, m^2 \frac{c_1 \left(\frac{R}{m^2}\right)^n}{c_2 \left(\frac{R}{m^2}\right)^n + 1},
\label{eq5}
\end{equation}
where $n>0$, $c_1$ and $c_2$ are dimensionless model parameters and 
$m$ is another parameter, the square of which depicts the mass (energy) scale 
\cite{Sawicki_2007}. The functional form of the Hu-Sawicki $f(R)$ model given 
by Eq.~\eqref{eq5} is such that it fulfills certain desirable observational 
criteria. Without invoking a true cosmological constant, it can drive 
late-time cosmic acceleration with an expansion history that closely 
resembles the $\Lambda$CDM model \cite{2024_kumar,2012_Santos}. It satisfies 
an asymptotic behavior similar to the $\Lambda$CDM model in the large 
curvature region. The model, at $R \to \infty$ yields $f(R)=$ constant, thus 
mimicking the $\Lambda$CDM model in the large curvature region. Moreover, at 
$R\to 0$ it gives $f(R)=0$ \cite{Dhruba_2020}. It is important to emphasize 
that, similar to other viable $f(R)$ models, the Hu-Sawicki $f(R)$ function is 
also designed to meet the stability requirement $d^2f(R)/dR^2>0$ 
\cite{Sawicki_2007, 2010_Felice,2024_Chakraborty}. Furthermore, its field 
equations can be rewritten in a second-order form, allowing the theory to be 
reformulated as a second-order scalar-tensor theory. This is achieved by 
identifying the extra degree of freedom $F$ as a scalar field $\phi$~\cite{Nashiba_2021}. It ensures that the theory remains free from the Ostr\"{o}gradsky instability. Additionally, the explicit dependence of $F$ on curvature 
$R$ dynamically couples the scalar field $\phi$ to $R$. Consequently, 
variations of curvature caused by matter distributions directly impact the 
scalar field. 

\begin{figure}[!b]
        \centerline{
                \includegraphics[scale = 0.726]{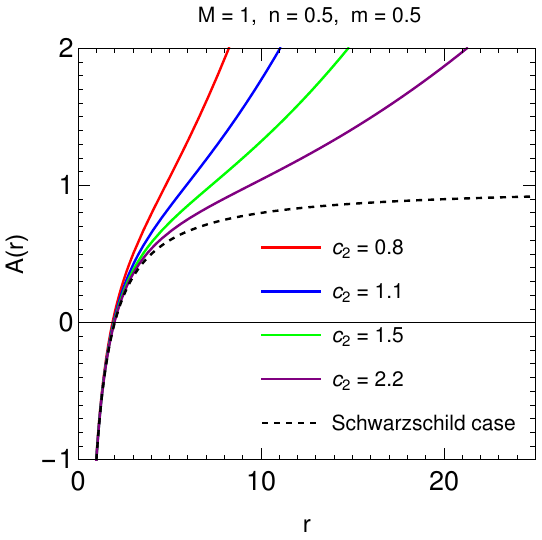}\hspace{1cm}
                \includegraphics[scale = 0.712]{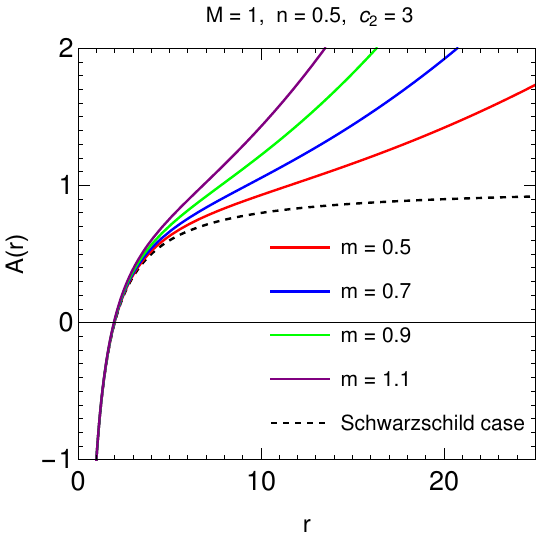}}\vspace{0.5cm}
        \centerline{
        	    \includegraphics[scale = 0.728]{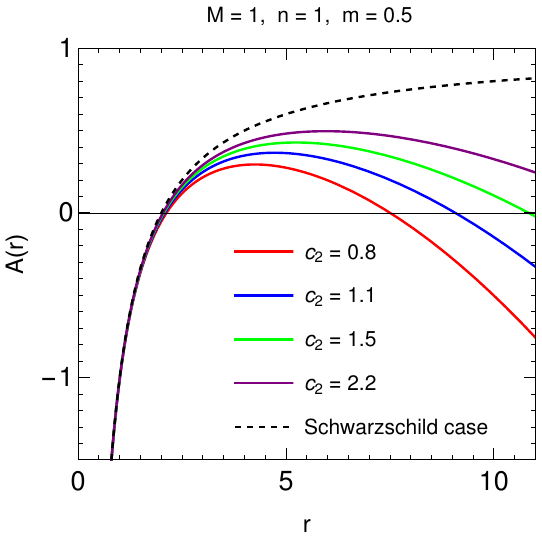}\hspace{1cm}
                \includegraphics[scale = 0.73]{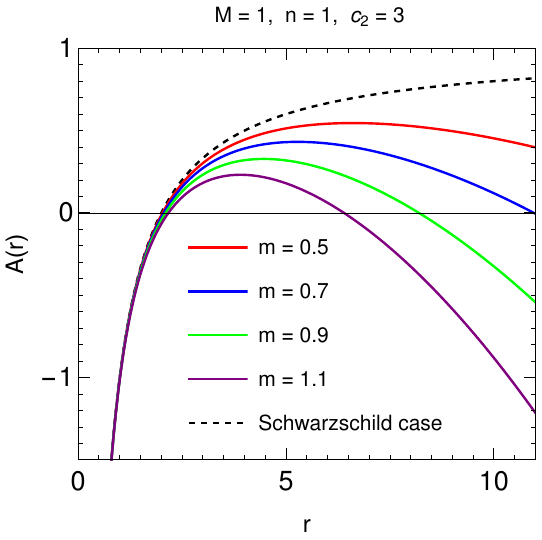}}
                \vspace{-0.2cm}
                \caption{The behavior of the metric function $A(r)$ with 
respect to radial distance $r$ for Hu-Sawicki model BHs
with different values of model parameters. The left two plots illustrate the 
behavior of the function as $c_2$ changes for $n = 0.5$ and $n = 1$ 
respectively. The right two plots display the behavior of the function with 
variation of $m$ for the same values of the parameter $n$.}
        \label{fig1}
\end{figure}
\textbf{Now}, the considered metric ansatz is given by
\begin{equation}
ds^2 = - A(r) dt^2 + B(r) dr^2 + r^2(d\theta^2 + \sin^2\!\theta\, d\phi^2).
\label{eq6}
\end{equation}
The metric coefficients of the spacetime that are deduced for the Hu-Sawicki 
$f(R)$ gravity model in the form \cite{2024_karm}:
\begin{align}
	A(r) = \frac{1}{B(r)} & = 1 - \frac{2M}{r} + \frac{m^2}{12} \left(\frac{n-2}{2c_2}\right)^{\!1/n}\!\!\! r^2 \nonumber\\[5pt]
         & = 1-\frac{2 M}{r} + \lambda\, r^2,
	\label{eq7}
\end{align}
where $M$ is the BH mass parameter and 
$\lambda = m^2/12\left((n-2)/2c_2\right)^{\!1/n}$.
It can be seen from the aforementioned relation that the BH spacetime 
does not depend on the model parameter $c_1$. Hence, in our work, we shall 
analyze the effect of the parameters $c_2$ and $m$ on the deflection angle of 
the BH spacetime in the weak as well as strong field regime. Also, we 
shall investigate the effect of these parameters on the lensing observables in 
the strong field limit. 

However, before proceeding to the proposed analyses, for completeness, 
it would be appropriate to make a few comments on the considered BH solution. 
It is seen from Eq.~\eqref{eq7} that as $\lambda$ equals to zero, this 
solution reduces to the Schwarzschild BH solution. The horizons of the BHs 
represented by Eq.~\eqref{eq7} are the positive real roots of equation 
$A(r)=0$. Depending on different sets of parameter values the solution reveals 
that the 
BHs exhibit either a single, well-defined event horizon analogous to the 
Schwarzschild BH or two distinct horizons as shown in Fig.~\ref{fig1}. It is 
observed that in the presence of two horizons, the outer horizon moves away 
from the inner one when $c_2$ increases whereas it approaches the inner horizon
as $m$ increases. Beyond a specific higher value of the parameter $m$ and a 
sufficiently lower value of $c_2$, the BHs become a horizonless singularity 
for the given values of other parameters.

\section{Gravitational bending of light in the weak field limit}
\label{sec.3}

A key parameter in probing the captivating phenomenon of gravitational lensing 
is the bending angle of light. In the weak field limit of deflection of light, 
the point of closest approach of photons traveling from a distant source is 
found to be far away from the lensing mass which in our case is the mass of the
BH. In this section, the deflection angle shall be deduced in the weak 
field limit of the Hu-Sawicki $f(R)$ gravity BH spacetime \eqref{eq6}. 
We tend to analyze the effect of the model parameters on the angle of bending 
of light as it approaches the weak gravitational field of the BH. With 
that in mind, we follow the Ishihara {\it et al.}~approach for asymptotically 
non-flat spacetimes presented in Ref.~\cite{Ishihara_2016}. 
\begin{figure}[ht!]
\centering
\includegraphics[scale=0.35]{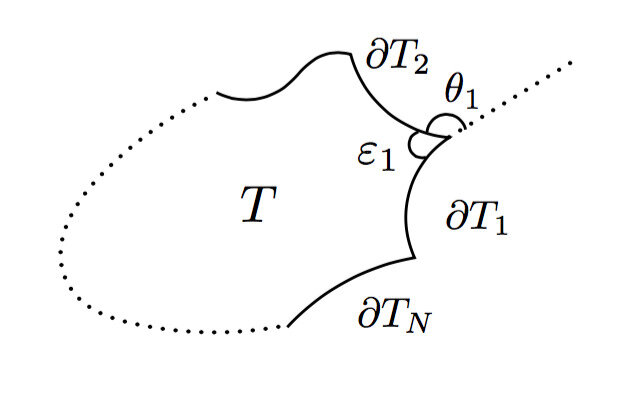}
\vspace{-0.5cm}
\caption{Schematic representation for the GBT \cite{Ishihara_2016}. The inner 
angle is $\varepsilon_a$ and the jump angle is $\theta_a$ ($a =1,2, ...,N$).}
\label{GL1}
\end{figure}
In this approach, the GBT is employed to compute the angle of bending of light. 
Several formulations of GBT exist and among them, the simplest one theorizes 
that the total Gaussian curvature surrounded by an enclosed triangle can be 
denoted in terms of the total geodesic curvature of the boundary and the jump 
angles formed at the corners. This theorem can be comprehended by following 
the illustration shown in Fig.~\ref{GL1}. An orientable surface $T$ is 
portrayed in two dimensions, whose boundaries are differentiable curves. These 
curves are represented as $\partial T_a$ ($a=1,2,...,N$) with $\theta_a$ as 
the jump angles formed between the curves. Consequently, the GBT can be 
mathematically defined as \cite{Carmo_2016}
\begin{equation}
\int \int_T \mathcal{K} dS + \sum^{N}_{a\,=\,1} \int_{\partial T_a} \kappa_g dl + \sum^{N}_{a\,=\,1} \theta_a = 2\pi,
\end{equation}
where $\mathcal{K}$ denotes the Gaussian curvature of the surface $T$, 
$\kappa_g$ is the geodesic curvature of the boundaries $\partial T_a$ with an 
infinitesimal line element $dl$ along the boundary. The sign of $dl$ is chosen 
in such a way that it is in accordance with the orientation of the surface 
following $dl>0$ for prograde motion and $dl<0$ for retrograde motion of 
photons.

\begin{figure}[ht!]
\centering
\includegraphics[scale=0.28]{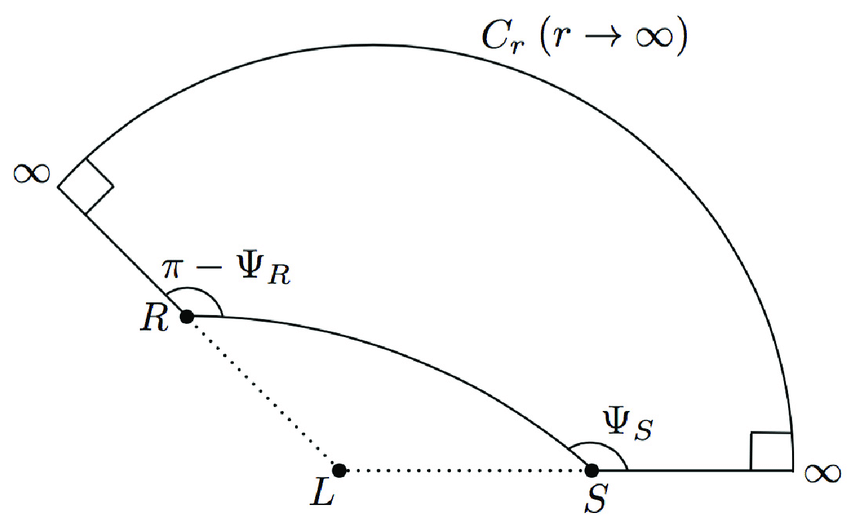}
\caption{Illustrative description for the quadrilateral $\stackrel{\infty}{R}\!\square\!\stackrel{\infty}{S}$ enclosed in a curved space \cite{Takizawa_2020}.}
 \label{GL2}
\end{figure}

In Fig.~\ref{GL2}, the BH is portrayed as a lens ($L$) with the source 
($S$) and the receiver ($R$) located at a finite distance from $L$. In view of 
the equatorial plane ($theta = \pi/2$), the deflection angle of light 
approaching from the source can be presented as \cite{Ishihara_2016, Ono_2017}
\begin{equation}
\hat{\alpha} = \Psi_R - \Psi_S + \phi_{RS},
\label{angle1}
\end{equation}
where $\Psi_R$ and $\Psi_S$ exhibit the angles of light approximated with 
respect to $L$ at the positions of $S$ and $R$ respectively. $\phi_{RS}$ is 
the separation angle between $R$ and $S$, and is signified by 
$\phi_{RS} = \phi_R - \phi_S$ where $\phi_R$ and $\phi_S$ are the longitudes 
of $R$ and $S$ respectively. 

The null condition which signifies $ds^2 = 0$ is followed by the light rays. 
Consequently, the BH metric can be recast as
\begin{equation}
dt^2 = \gamma_{ij} dx^i dx^j = \frac{1}{A(r)^2}\, dr^2 + \frac{r^2}{A(r)}\, d\Omega^2,
\label{nullgeod}
\end{equation}
where $\gamma_{ij}$ is usually indicated as the optical metric and 
$d\Omega^2 = d\theta^2 + \sin^2\!\theta\,d\phi^2$. It describes 
a 3D Riemannian space represented by $\mathcal{M}^{(3)}$. In this manifold, a 
ray of light is interpreted as a spatial curve. The non-vanishing components 
of the optical metric are
\begin{equation}
\gamma_{rr} = \frac{1}{A(r)^2},\;\; \gamma_{\phi\phi} = \frac{r^2}{A(r)}.
\label{compt}
\end{equation}

Another parameter of paramount importance in the study of the bending of light 
is the impact parameter. It is typically defined as the ratio of the angular 
momentum ($\mathcal{L}$) and the energy ($\mathcal{E}$) of photons. In the 
equatorial plane of spacetime, $\mathcal{L}$ and $\mathcal{E}$ are the 
constants of motion and for the spacetime~\eqref{eq6}, these can be 
represented as $\mathcal{E} = A(r)\,\dot{t}$ and 
$\mathcal{L} = r^2\,\dot{\phi}$, where the over dot depicts the derivative 
with respect to the affine parameter $\tau$ along the path of the light ray. 
Hence, the impact parameter is given as 
\begin{equation}
\zeta \equiv \frac{\mathcal{L}}{\mathcal{E}} = \frac{r^2}{A(r)} \frac{d\phi}{dt}.
\label{impact}
\end{equation}
The unit radial vector along the radial direction from the center of the lens 
and the unit angular vector across the angular path are respectively obtained 
as $e_{rad} = (A(r), 0)$ and $e_{ang} = (0, A(r)/r)$. Moreover, the 
components of the unit tangent vector $\bm{K} \equiv d\bm{x}/dt$ along the
path of the light ray are obtained as \cite{Ishihara_2016}
\begin{equation}
(K^r , K^\phi) = \frac{\zeta A(r)}{r^2} \left(\frac{dr}{d\phi} , 1\right).
\label{eq13}
\end{equation}
The term $dr/d\phi$ in the above expression results in the orbit equation as  
\begin{equation}
\left(\frac{dr}{d\phi}\right)^{\!2} = -\, r^2 A(r) + \frac{r^4}{\zeta^2}.
\label{orbiteqn}
\end{equation}
Now, if $\Psi$ is assumed to be the angle of the light ray estimated from the 
radial direction, then we have
\begin{equation}
\cos\Psi = \frac{\zeta}{r^2} \frac{dr}{d\phi},
\label{eq15}
\end{equation}
which leads to,
\begin{equation}
\sin\Psi = \frac{\zeta \sqrt{A(r)}}{r}.
\label{eq16}
\end{equation}
Again, considering a new variable $u = 1/r$, we can recast 
Eq.~\eqref{orbiteqn} as 
\begin{equation}
\left(\frac{du}{d\phi}\right)^2 = F(u),
\label{eq17}
\end{equation}
where the function $F(u)$ is obtained as $F(u) = -\,u^2\,A(u) + 1/\zeta^2$.

At this stage, it needs to be pointed out that the quadrilateral 
$\stackrel{\infty}{R}\!\square\!\stackrel{\infty}{S}$ portrayed in 
Fig.~\ref{GL2} is enclosed within the space $\mathcal{M}^{(3)}$. This 
quadrilateral $\stackrel{\infty}{R}\!\square\!\stackrel{\infty}{S}$ comprises 
of light rays behaving as spatial curves with two outgoing radial lines from 
receiver and source, along with a circular arc fragment $C_r$ with the 
coordinate radius $r_C$ ($r_C \rightarrow \infty$). It is clearly observed 
from Fig.~\ref{GL2} that within the asymptotically flat Minkowskian spacetime, 
$\kappa_g \rightarrow 1/r_C$ and $dl \rightarrow r_C\, d\phi$ as 
$r_C \rightarrow \infty$ \cite{Gibbons_2008}. Accordingly, we can present the 
bending angle of light in the domain 
$\stackrel{\infty}{R}\!\square\!\stackrel{\infty}{S}$ as
\begin{equation}
\hat{\alpha} = \Psi_R - \Psi_S + \phi_{RS} = - \int \int_{\stackrel{\infty}{R}\square\stackrel{\infty}{S}} \mathcal{K}\, dS. 
\label{eq18}
\end{equation}
Integrating Eq.~\eqref{eq17}, the separation angle $\phi_{RS}$ can be obtained 
as
\begin{equation}
\phi_{RS} = 2\int_{0}^{u_0}\!\!\!\frac{du}{\sqrt{F(u)}},
\label{eq19}
\end{equation}  
where $u_0$ denotes the inverse of the distance of the closest approach. 
Corresponding to the Ishihara $\textit{et al.}$ approach followed in this 
work, if $S$ and $R$ positions are considered to be at finite distances from 
the BH, the gravitational deflection angle can be represented as 
\begin{equation}
\hat{\alpha} = \Psi_R - \Psi_S + \int_{u_R}^{u_0}\!\!\! \frac{du}{\sqrt{F(u)}} 
+ \int_{u_S}^{u_0}\!\!\! \frac{du}{\sqrt{F(u)}}. 
\label{eq20}
\end{equation}

Now, applying Eq.~\eqref{eq16} to the BH metric \eqref{eq6}, we 
arrive at
\begin{equation}
\begin{split}
\hspace{0.5cm}\Psi_R - \Psi_S &= \arcsin(\zeta u_R) + \arcsin(\zeta u_s) - \pi + \zeta M \left(\frac{u_R^2}{\sqrt{1 - \zeta^2 u_R^2}} + \frac{u_S^2}{\sqrt{1 - \zeta^2 u_S^2}}\right) \\
&+ \frac{2^{-2 - \frac{1}{n}}}{3}\zeta^3 m^2 M \left(\frac{u_R^2}{(1 - \zeta^2 u_R^2)^{3/2}} + \frac{u_S^2}{(1 - \zeta^2 u_S^2)^{3/2}}\right) + \frac{\zeta M^2}{2} \left(\frac{u_R^3}{(1 - \zeta^2 u_R^2)^{3/2}} + \frac{u_S^3}{(1 - \zeta^2 u_S^2)^{3/2}}\right) \\ 
&- \frac{2^{-3-\frac{1}{n}}}{3}\zeta m^2 M \left(\frac{1}{(1 - \zeta^2 u_R^2)^{3/2}} + \frac{1}{(1 - \zeta^2 u_S^2)^{3/2}}\right) - 2^{-4-\frac{1}{n}}\zeta m^2 M^2\left(\frac{n - 2}{c_2}\right)^{1/n}\left(\frac{u_R}{(1 - \zeta^2 u_R^2)^{5/2}}\right. \\
&\left. + \frac{u_S}{(1 - \zeta^2 u_S^2)^{5/2}}\right) + \frac{2^{-1-\frac{1}{n}}}{3}\zeta^3 m^2 M^2 \left(\frac{n-2}{c_2}\right)^{1/n} \left(\frac{u_R^3}{(1 - \zeta^2 u_R^2)^{5/2}} + \frac{u_S^3}{(1 - \zeta^2 u_S^2)^{5/2}}\right) \\
&- \frac{2^{-3-\frac{1}{n}}}{3} \zeta m^2 \left(\frac{n-2}{c^2}\right)^{1/n} \left(\frac{u_R^{-1}}{\sqrt{1 - \zeta^2 u_R^2}} + \frac{u_S^{-1}}{\sqrt{1 - \zeta^2 u_S^2}}\right).
\end{split}
\label{eq21}
\end{equation}
It is evident that this expression tends to become divergent at 
$u_R \rightarrow 0$ and $u_S \rightarrow 0$ due to the fact that the 
spacetime under consideration is asymptotically non-flat. Thus, this series 
Eq.~\eqref{eq21} must be employed only within a certain limit of the finite 
radius of convergence.

Next, the angle $\phi_{RS}$ is computed for the BH spacetime \eqref{eq6} as
\begin{equation}
\begin{split}
\phi_{RS} &= \pi - \arcsin(\eta u_R) - \arcsin(\eta u_S) + \left[\frac{15M^2}{4\zeta^2} + 5\times 2^{-4-\frac{1}{n}}m^2 M^2 \left(\frac{n-2}{c_2}\right)^{1/n}\right] \left(\pi - \arcsin(\eta u_R) \right.\\[5pt]
&\left. - \arcsin(\eta u_S) \right) - \frac{2^{-3-\frac{1}{n}}}{3} \zeta^3 m^2 \left(\frac{n-2}{c_2}\right)^{1/n} \left(\frac{u_R}{\sqrt{1 - \zeta^2 u_R^2}} + \frac{u_S}{\sqrt{1 - \zeta^2 u_S^2}}\right) \\[5pt]
&+ \left[\frac{2M}{\zeta} - \frac{2^{-2-\frac{1}{n}}}{3} \zeta m^2 M \left(\frac{n-2}{c_2}\right)^{1/n}\right]\left(\frac{1}{(1 - \zeta^2 u_R^2)^{3/2}} + \frac{1}{(1 - \zeta^2 u_S^2)^{3/2}}\right) \\[5pt]
&- \left[3\zeta M - 2^{-2-\frac{1}{n}} \zeta^3 m^2 M \left(\frac{n-2}{c_2}\right)^{1/n}\right] \left(\frac{u_R^2}{(1 - \zeta^2 u_R^2)^{3/2}} + \frac{u_S^2}{(1 - \zeta^2 u_S^2)^{3/2}}\right) \\[5pt]
&+ \left[\frac{15 M^2}{4\zeta} + 5\times 2^{-4-\frac{1}{n}}\zeta m^2 M^2 \left(\frac{n-2}{c_2}\right)^{1/n}\right] \left(\frac{u_R^2}{(1 - \zeta^2 u_R^2)^{5/2}} + \frac{u_S^2}{(1 - \zeta^2 u_S^2)^{5/2}}\right) \\[5pt]
&- \left[\frac{35\zeta M^2}{4} + \frac{35\times 2^{-4-\frac{1}{n}}}{3} \zeta^3 m^2 M^2 \left(\frac{n-2}{c_2}\right)^{1/n}\right] \left(\frac{u_R^3}{(1 - \zeta^2 u_R^2)^{5/2}} + \frac{u_S^3}{(1 - \zeta^2 u_S^2)^{5/2}}\right).
\end{split}
\label{eq22}
\end{equation}
Finally, by combining Eqs.~\eqref{eq21} and ~\eqref{eq22}, we arrive at the 
bending angle of light in the gravitational field of the Hu-Sawicki $f(R)$ 
gravity BH under the study and is presented as
\begin{equation}
\begin{split}
\hat{\alpha} &= \left[\frac{15M^2}{4\zeta^2} + 5\times 2^{-4-\frac{1}{n}}m^2 M^2 \left(\frac{n-2}{c_2}\right)^{1/n}\right] \left(\pi - \arcsin(\eta u_R) - \arcsin(\eta u_S)\right) \\[5pt]
&- 2^{-2-\frac{1}{n}} \zeta^3 m^2 \left(\frac{n-2}{c_2}\right)^{1/n} \left(\frac{u_R}{\sqrt{1 - \zeta^2 u_R^2}} + \frac{u_S}{\sqrt{1 - \zeta^2 u_S^2}}\right) + \zeta M \left(\frac{u_R^2}{\sqrt{1 - \zeta^2 u_R^2}} + \frac{u_S^2}{\sqrt{1 - \zeta^2 u_S^2}}\right) \\[5pt]
&+ \left[3\times 2^{-3-\frac{1}{n}} \zeta^3 m^2 M \left(\frac{n-2}{c_2}\right)^{1/n} - 3\zeta M\right] \left(\frac{u_R^2}{(1 - \zeta^2 u_R^2)^{3/2}} + \frac{u_S^2}{(1 - \zeta^2 u_S^2)^{3/2}}\right) \\[5pt]
&+ \frac{\zeta M^2}{2} \left(\frac{u_R^3}{(1 - \zeta^2 u_R^2)^{3/2}} + \frac{u_S^3}{(1 - \zeta^2 u_S^2)^{3/2}}\right) + \left[\frac{2M}{\zeta} - 2^{-3-\frac{1}{n}} \zeta m^2 M \left(\frac{n-2}{c_2}\right)^{1/n}\right] \\[5pt]
& \times \left(\frac{1}{(1 - \zeta^2 u_R^2)^{3/2}} + \frac{1}{(1 - \zeta^2 u_S^2)^{3/2}}\right) - 2^{-4-\frac{1}{n}} \zeta m^2 M \left(\frac{n-2}{c_2}\right)^{1/n} \left(\frac{u_R}{(1 - \zeta^2 u_R^2)^{5/2}} + \frac{u_S}{(1 - \zeta^2 u_S^2)^{5/2}}\right) \\[5pt]
&+ \left[\frac{15M^2}{4\zeta} + 5\times 2^{-4-\frac{1}{n}} \zeta m^2 M^2 \left(\frac{n-2}{c_2}\right)^{1/n}\right] \left(\frac{u_R^2}{(1 - \zeta^2 u_R^2)^{5/2}} + \frac{u_S^2}{(1 - \zeta^2 u_S^2)^{5/2}}\right) \\[5pt]
&+ \left[- \frac{35\zeta M^2}{4} + 9\times 2^{-4-\frac{1}{n}} \zeta^3 m^2 M^2 \left(\frac{n-2}{c_2}\right)^{1/n}\right] \left(\frac{u_R^3}{(1 - \zeta^2 u_R^2)^{5/2}} + \frac{u_S^3}{(1 - \zeta^2 u_S^2)^{5/2}}\right) \\[5pt]
&- \frac{2^{-3-\frac{1}{n}}}{3} \zeta m^2 \left(\frac{n-2}{c_2}\right)^{1/n} \left(\frac{u_R^{-1}}{\sqrt{1 - \zeta^2 u_R^2}} + \frac{u_S^{-1}}{\sqrt{1 - \zeta^2 u_S^2}}\right).
\end{split}
\label{angle1}
\end{equation}
As a consequence of a few terms present in Eq.~\eqref{eq21}, the above 
expression also becomes divergent in the far distance limit 
$(u_R \rightarrow 0, u_S \rightarrow 0)$. The reason behind this was stated 
earlier as that the spacetime under study is asymptotically 
non-flat, analogous to the Kottler spacetime \cite{Kottler_1918} employed by 
Ishihara {\textit{et al.}}~\cite{Ishihara_2016}. In their work, Ishihara 
{\textit{et al.}}~discussed this issue of the divergence of the 
deflection angle and stated that such an issue is not of much concern as the 
limit $u_R \rightarrow 0$, $u_S \rightarrow 0$ is not applicable for 
astronomical observations. Also, due to the analogy between the BH 
metric~\eqref{eq7} and the Kottler spacetime mentioned above, the deflection 
angle derived in our study appears to coincide with that derived in 
Ref.~\cite{Ishihara_2016}. However, for both instances, the effective 
cosmological constant will affect the deflection angle in significantly 
different manners. The reason behind this is that in our case, the effective 
cosmological constant depends on two important Hu-Sawicki model parameters 
$m$ and $c_2$. Furthermore, the effect of the Hu-Sawicki model parameter on 
the deflection angle $\hat{\alpha}$ can be evidently analyzed 
from the aforementioned equation. However, if the model parameters are to 
vanish, i.e.~$n = m = c_2 = 0$, it would result in
 
\begin{equation}
\begin{split}
\hat{\alpha} &= \frac{15M^2}{4\zeta^2} \left(\pi - \arcsin(\eta u_R) - \arcsin(\eta u_S)\right) + \zeta M \left(\frac{u_R^2}{\sqrt{1 - \zeta^2 u_R^2}} + \frac{u_S^2}{\sqrt{1 - \zeta^2 u_S^2}}\right) \\[5pt]
&- 3\zeta M \left(\frac{u_R^2}{(1 - \zeta^2 u_R^2)^{3/2}} + \frac{u_S^2}{(1 - \zeta^2 u_S^2)^{3/2}}\right) + \frac{\zeta M^2}{2} \left(\frac{u_R^3}{(1 - \zeta^2 u_R^2)^{3/2}} + \frac{u_S^3}{(1 - \zeta^2 u_S^2)^{3/2}}\right) \\[5pt]
&+ \frac{2M}{\zeta} \left(\frac{1}{(1 - \zeta^2 u_R^2)^{3/2}} + \frac{1}{(1 - \zeta^2 u_S^2)^{3/2}}\right) + \frac{15M^2}{4\zeta} \left(\frac{u_R^2}{(1 - \zeta^2 u_R^2)^{5/2}} + \frac{u_S^2}{(1 - \zeta^2 u_S^2)^{5/2}}\right) \\[5pt]
&- \frac{35\zeta M^2}{4} \left(\frac{u_R^3}{(1 - \zeta^2 u_R^2)^{5/2}} + \frac{u_S^3}{(1 - \zeta^2 u_S^2)^{5/2}}\right). 
\end{split}
\label{angle2}
\end{equation}
which in the far distance limit $(u_R \rightarrow 0, u_S \rightarrow 0)$ yields the bending angle of light for a Schwarzschild BH and is deduced as
\begin{equation}
    \hat{\alpha} \simeq  \frac{4M}{\zeta} + \frac{15M^2\pi}{4\zeta^2}
\label{schwz}
\end{equation}

\begin{figure}[!htp]
   \centering
{\includegraphics[width=0.423\textwidth]{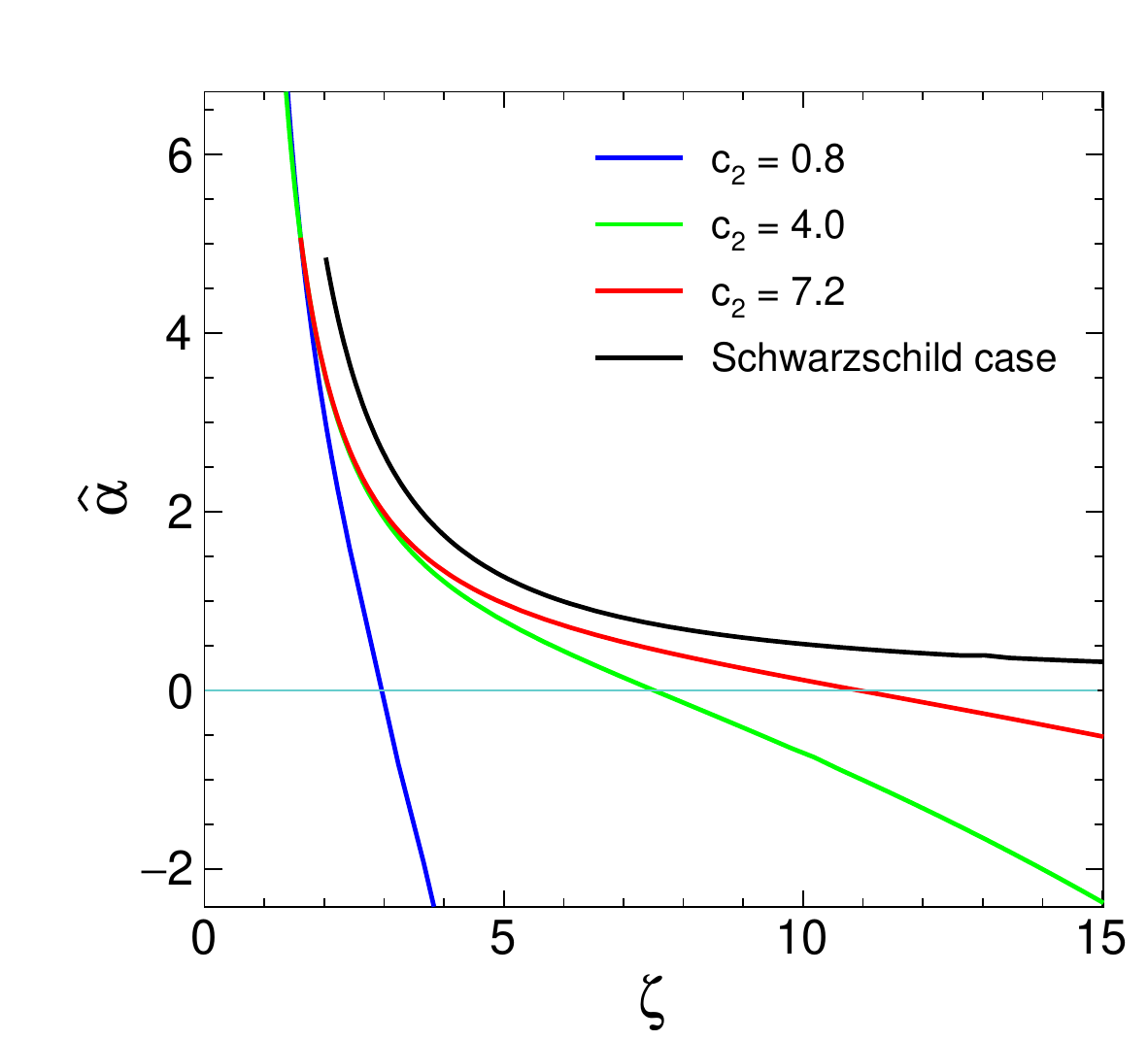}
\label{Deflect1}}\hspace{0.5cm} 
{\includegraphics[width=0.41\textwidth]{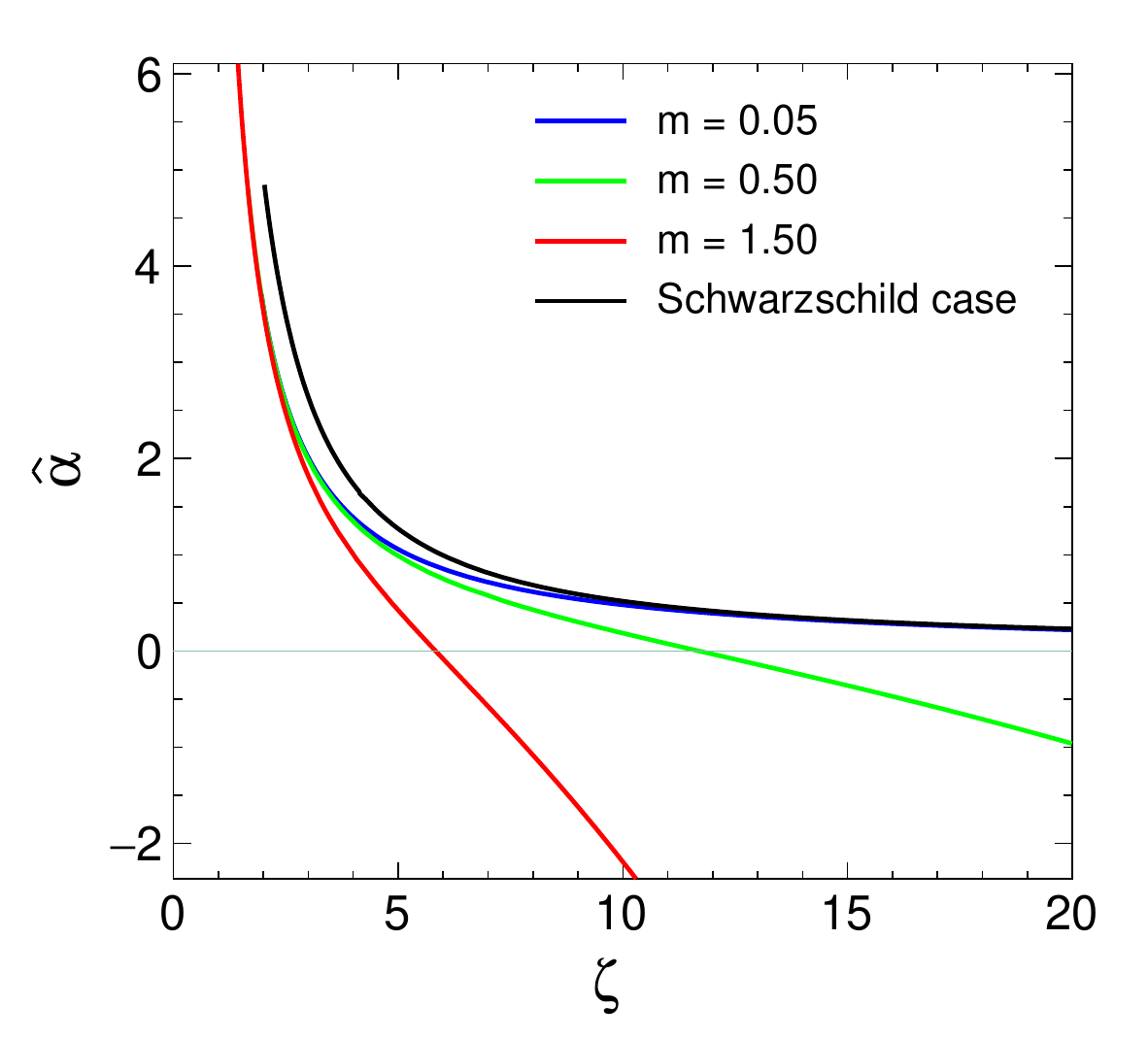}
\label{Deflect2}}  
\caption{Deflection angle as a function of the impact parameter $\zeta$ for 
different values of the Hu-Sawicki model parameters $c_2$ and $m$ with $M=1$, 
and $n=0.5$.}
\label{Deflection1}
\end{figure}

In Fig.~\ref{Deflection1}, we portray the deflection angle formed as a result 
of the bending of light in the weak gravitational field of the Hu-Sawicki 
$f(R)$ gravity BH as a function of the impact parameter and analyze 
the effects of the model parameters $m$ and $c_2$ on the deflection angle. In 
our analysis, we have taken $u_R = u_S = 0.5/\zeta$. It is evident from the 
two graphical depictions displayed in Fig.~\ref{Deflection1} that for each 
plot the computed results are compared with that of the Schwarzschild case. 
We can refer from each representation that for both cases, the deflection 
angle decreases with an increase in the impact parameter analogous to the 
Schwarzschild case up to a certain limit and then unlike the Schwarzschild 
case, the deflection angle continues to decrease with increasing impact 
parameter and becomes negative. In the first illustration, we have considered 
three different values of $c_2 = 0.8, 4.0, 7.2$ and compared the deflection 
angle for these values with the Schwarzschild case. It is observed from the 
figure that as the value of $c_2$ increases, the deflection angle tries to 
mimic the Schwarzschild behaviour. However, when observed for much higher 
impact parameters , the deflection angle is still found to stay on the negative 
end. For the case of different $m$ values $m = 0.05, 0.5, 1.50$, it can be 
seen that for lower $m$ values, the deflection angle slowly moves towards the 
positive end for higher impact parameters. For $m = 0.05$, the deflection 
angle is seen to overlap with the Schwarzschild case for higher impact 
parameter values. Thus, it can be said that for smaller $m$ values, the 
Schwazschild behaviour can be recovered from the Hu-Sawicki $f(R)$ gravity 
BH beyond certain values of the impact parameter. For the negative 
deflection angle, it can be remarked that for high values of the impact 
parameter, the photons suffer repulsion by the gravitational field of the 
BH. Such a negative deflection angle presents an idea regarding the 
nature of the gravitational field of the BH. A number of research 
works \cite{Nashiba_2023c, Panpanich_2019, Nakashi_2019, Kitamura_2013} have 
also arrived at a negative deflection angle.
 
\section{Gravitational lensing in the strong field limit}
\label{sec.4}
In the strong field gravitational lensing, a photon from a distant
source approaches very close to a massive object such as a BH and experiences
an intense gravitational field of the object. Consequently, it suffers
significant deviation from its path with increasing deflection angle for
decreasing distance from the BH. In this section, we shall determine the
deflection angle and lensing observables in the strong field regime for a 
static, spherically symmetric BH as described by the line element 
\eqref{eq6}~\cite{1972_stefan,2002_Bozza,2017_Naoki}, where the metric 
function $A(r)$ and $B(r)$ are connected as given by Eq.~\eqref{eq7} and 
$C(r) = r^2$. Also, we will evaluate here the lensing observables for a 
few known supermassive BHs. It should be noted that as in the weak field limit, 
for this analysis also we confine the photon's entire trajectory in the 
equatorial plane ($\theta=\pi/2$) only.

\subsection{Deflection angle}
Both the deflection angle and lens equation are key factors for understanding 
the behavior of strong field gravitational lensing. In this scenario, a photon 
originating from a distant source and possessing a specific impact parameter 
is deflected by the BH as it reaches the closest approach distance $r_0$,  
which represents the turning point of its trajectory. Then it proceeds to the 
far-away observer in another direction 
\cite{2002_Bozza,2006_kabita,2023_Kumar,2024_islum}. As $r_0$ decreases, the 
deflection angle increases significantly. Once it reaches $2\pi$, the photon 
makes a loop entirely around the BH before arriving at the observer 
\cite{2002_Bozza,2006_kabita}. Further decrease of $r_0$, the deflection angle
exceeds $2\pi$ and the photon makes multiple loops around the BH before being 
escaped to infinity \cite{2002_Bozza,2020_islum}. When $r_0$ approaches the 
photon sphere radius $r_p$ the deflection angle diverges. For $r_0<r_p$, the 
incoming photon gets trapped by the BH and cannot come out from it. 
However, one can rewrite the trajectory of the photon given by 
Eq.~\eqref{orbiteqn} in terms of radial effective potential $V_{eff}(r)$ as
\begin{equation} 
	\frac{V_{eff}(r)}{E^2} = \frac{\zeta^2}{r^2}\,\left[1-\frac{2M}{r}+ \frac{m^2}{12} \left(\frac{n-2}{2c_2}\right)^{\!1/n}\!\!\! r^2\right].
	\label{eq29}
\end{equation} 
Eq.~\eqref{eq29} determines different types of photon orbits namely, orbits 
with $\zeta<\zeta_p$, $\zeta=\zeta_p$ and $\zeta >\zeta_p$, where $\zeta_p$ is 
the impact parameter which is the critical or minimum impact parameter 
evaluated at the photon sphere radius $r_p$. A photon with an impact parameter 
smaller than $\zeta_p$ falls into the event horizon, whereas one with an 
impact parameter beyond the critical value $\zeta_p$ is scattered by the BH 
towards a distant observer. On the other hand, $\zeta=\zeta_p$ corresponds to 
a photon that follows a circular path of constant radius $r_p$, which becomes 
unstable by a small change in its radial position \cite{2020_islum}. The 
potential of a photon sphere orbit is flat and satisfies the condition
$V^\prime_{eff}(r)=0$. This condition leads to the photon sphere equation  
\cite{2002_Bozza,2020_islum,Virbhadra_2002} as 
\begin{equation} 
	r A^\prime(r) - 2 A(r) =0.
	\label{eq30}
\end{equation}
In fact, the photon sphere radius is the greatest positive root of 
Eq.~\eqref{eq30} and in our case, it is found as 3$M$ for the BH solution 
\eqref{eq7}. Further, at $r=r_0$, the equation of orbit \eqref{orbiteqn} yields 
the relation between the closest approach distance and the impact parameter 
as $\zeta_0=\sqrt{C(r_0)/A(r_0)}$. From this, the critical impact parameter 
at $r_0=r_p$ can be defined as \cite{2002_Bozza,2020_islum}
 \begin{equation} 
 	\zeta_p=\sqrt{\frac{C(r_p)}{A(r_p)}},
 	\label{eq31}
 \end{equation}
which for the BH solution \eqref{eq7} takes the form:
\begin{equation} 
	\zeta_p=\frac{r_p}{\sqrt{1 - \frac{2M}{r_p} + \frac{m^2}{12} \left(\frac{n-2}{2c_2}\right)^{\!1/n}\!\!\! r_p^2}}.
	\label{eq32}
\end{equation}
In strong deflection limit, the deviation suffered by a light ray is 
characterized by the deflection angle \cite{2002_Bozza,1972_stefan,2022_kumar,2011_Eiroa} 
 \begin{equation} 
	\hat{\alpha}(r_0)=I(r_0) - \pi,  
	\label{eq33}
\end{equation}
where 
\begin{equation} 
I(r_0)= 2\!\int_{r_0}^{\infty}\!\frac{d\varphi}{dr}\, dr,\;\;\; \frac{d\varphi}{dr} = \frac{\sqrt{B(r)}}{\sqrt{C(r)}\sqrt{\frac{C(r)A(r_0)}{C(r_0)A(r)}-1}}.
\label{eq34}
\end{equation}

\begin{figure}[!h]
	\centerline{
		\includegraphics[scale = 0.72]{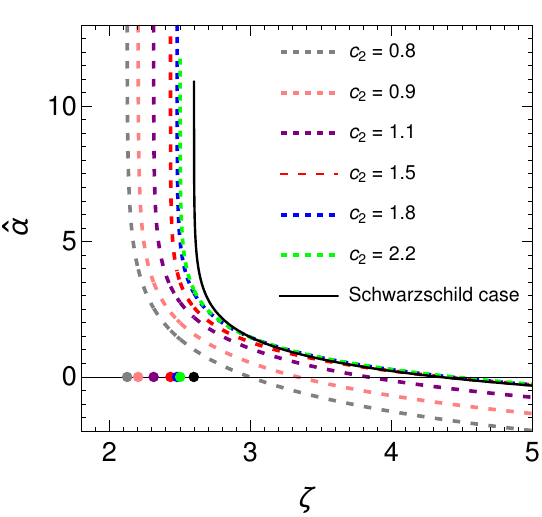}\hspace{0.7cm}
		\includegraphics[scale = 0.72]{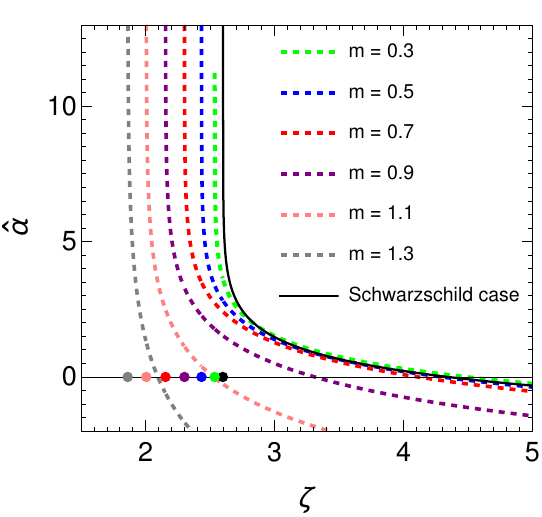}}\vspace{0cm}
\caption{Variation of deflection angle with impact parameter $\zeta$ for 
different values of the Hu-Sawicki model parameters where the points on the 
horizontal axis represent the value of impact parameter $\zeta = \zeta_p$ at 
which the deflection angle diverges. The model parameters $m = 0.5$ and 
$c_2 = 1.5$ are considered for the left and right panels respectively with 
$M=1$ and $n=0.5$ in both cases.}
\label{fig3} 
\end{figure}

To explore the behavior of the deflection angle we use the method developed by 
V.~Bozza, applicable to light rays governed by a standard geodesic equation in 
any spacetime and under any gravitational theory~\cite{2002_Bozza}. He has 
evaluated the diverging deflection angle at $r_0=r_p$ corresponding to an 
impact parameter $\zeta=\zeta_p$, by introducing a new variable $z$ given as
\begin{equation} 
z=\frac{A(r)-A(r_0)}{1-A(r_0)}.
	\label{eq36}
\end{equation}
Using this variable, integral $I(r_0)$ can be expressed 
as~\cite{2002_Bozza,2020_islum,Kuang_2022,2021_Gao} 
\begin{equation} 
	I(r_0)=\int_{0}^1\! H(z,r_0)\, g(z,r_0)\, dz,
	\label{eq37}
\end{equation}
where the functions $H(z,r_0)$ and $g(z,r_0)$ are defined as~\cite{2002_Bozza} 
\begin{equation} 
	 H(z,r_0)= \frac{2\sqrt{C(r_0)}\ (1-A(r_0))\ }{C(r)\,A^{'}(r_0)},
	\label{eq38}
\end{equation}
and
\begin{equation} 
	g(z,r_0)= \frac{1}{\sqrt{A(r_0)- \frac{A(r)}{C(r)}\,C(r_0)}}.
	\label{eq39}
\end{equation}
The function $H(z,r_0)$ is finite for all values of $z$ and $r_0$, whereas 
$g(z,r_0)$ exhibits divergence as $r_0$ approaches the radius of the photon 
sphere, $r_p$. To estimate the rate of this divergence, the function within 
the square root in $g(z,r_0)$ is expanded to second-order in $z$ that 
leads \cite{2002_Bozza,2020_islum} the following relation:
\begin{equation} 
	g(z,r_0)\sim g_0(z,r_0)= \frac{1}{\sqrt{\psi(r_0)\, z+ \eta(r_0)\, z^2}},
	\label{eq40}
\end{equation}
with
\begin{align} 
\psi(r_0) & = \frac{2 \left(1-\frac{3 M}{r_0}\right) \left(\frac{2 M}{r_0}-\frac{m^2}{12} \left(\frac{n-2}{2c_2}\right)^{\!1/n}\!\!\! r_0^2\right)}{\frac{2 M}{r_0} + \frac{m^2}{6} \left(\frac{n-2}{2c_2}\right)^{\!1/n}\!\!\! r_0^2},
	\label{eq41}\\[8pt]
\eta(r_0) & = \frac{\left(\frac{2 M}{r_0} - \frac{m^2}{12} \left(\frac{n-2}{2c_2}\right)^{\!1/n}\!\!\! r_0^2\right)^{\!2}}{2\,r_0^4 \left(\frac{2 M}{r_0^2} + \frac{m^2}{6} \left(\frac{n-2}{2c_2}\right)^{\!1/n}\!\!\! r_0\right)^{\!3}} \Bigg[4\, r_0^3 \left(\frac{2 M}{r_0^2} + \frac{m^2}{6} \left(\frac{n-2}{2c_2}\right)^{\!1/n}\!\!\! r_0 \right)^{\!2}\notag \\[5pt]
        &\quad  - \left(1-\frac{2 M}{r_0} + \frac{m^2}{12} \left(\frac{n-2}{2c_2}\right)^{\!1/n}\!\!\! r_0^2  \right) \left( 4 M +  \frac{4m^2}{3} \left(\frac{n-2}{2c_2}\right)^{\!1/n}\!\!\! r_0^3 \right) \Bigg].
\label{eq42}
\end{align}
Also, the variable $z$ in Hu-Sawicki model BH takes the form:
\begin{equation} 
	z = {2M}\left(\frac{r-r_0}{rr_0}\right) + \frac{m^2}{12} \left(\frac{n-2}{2c_2}\right)^{\!1/n}\!\!\! \left(r^2- r_0^2\right).
	\label{eq43}
\end{equation}
At $r_0=r_p$, the coefficient $\psi(r_0)$ of $z$ becomes zero that makes 
$g_0(z,r_0)$ to behave as $z^{-1}$ resulting a logarithmic divergence of the 
integral~\eqref{eq37}. Moreover, the integral $I(r_0)$ can be solved by 
dividing it into the divergent and regular terms as~\cite{2002_Bozza}
\begin{equation} 
	I(r_0)= I_D(r_0) + I_R (r_0),
	\label{eq44}
\end{equation}
where
\begin{equation} 
	I_D(r_0)=\int_{0}^1\!H(0,r_p)\, g_0(z,r_0)\,dz,
	\label{eq45}
\end{equation}
\begin{equation} 
	I_R(r_0)=\int_{0}^1\!\big[H(z,r_0)\, g(z,r_0) -H(0,r_p)\, g_0(z,r_0)\big]\,dz.
	\label{eq46}
\end{equation}
Further analysis of integrals~\eqref{eq45} and~\eqref{eq46} in strong field 
limit and approximating them using the leading terms an analytical expression 
for the deflection angle in terms of impact parameter near the point of 
divergence can be established in the following form~\cite{2002_Bozza}: 
\begin{equation}
\hat{\alpha}(\zeta)=-\,\bar{a}\log \left(\frac{\zeta}{\zeta_p}-1\right) + \bar{b} + \mathcal{O} (\zeta-\zeta_p),
\label{eq47}
\end{equation}
where the strong deflection coefficients $\bar {a}$ and $\bar{b}$ are given as
\begin{align} 
	\bar{a} & = \frac{a}{2} = \frac{H(0,r_p)}{2\, \sqrt{\eta(r_p)}},
	\label{eq48}\\[8pt]
	\bar{b} & = -\,\pi + b_R + b_D, 
	\label{eq49}
\end{align}
where
\begin{align}
b_D & =\bar{a}\, \log \Bigg[2\, \eta(r_p)\left(\frac{m^2}{12}\, r_p^2  \left(\frac{n-2}{2 c_2}\right)^{\!1/n}\!\!\!-\frac{2 M}{r_p}+1\right)^{\!-\,1}\Bigg], 
	\label{eq50}\\[8pt]
	b_R & = I_R(r_p) = \int_{0}^1\!\big[H(z,r_p)\, g(z,r_p) -H(0,r_p)\, g_0(z,r_p)\big]\,dz,
	\label{eq51}
\end{align}
and $\zeta_p = \zeta(r_p)$ as given by Eq.~\eqref{eq32}. We proceed by 
utilizing Eq.~\eqref{eq47} to obtain the strong deflection angle for the BH 
under consideration. To do this, we first calculate $\bar{a}$ employing 
Eq.~\eqref{eq48}, followed by computing $b_D$ based on $\eta(r_p)$ and 
$\bar{a}$ as specified in Eqs.~\eqref{eq42} and \eqref{eq48} respectively at 
$r_0 =r_p$. The integral $b_R$ represented by Eq.~\eqref{eq51} is estimated 
numerically where Eqs.~\eqref{eq39} and \eqref{eq40} are used along with 
Eq.~\eqref{eq36}. Employing all these values into Eq.~\eqref{eq47} we obtain 
the deflection angle for our BH \eqref{eq7}. To examine the behavior of the 
deflection angle, we depict its variation with impact parameter $\zeta$ for 
different values of model parameters with $M = 1$ in Fig.~\ref{fig3}, where 
the left plot is for different values $c_2$ fixing $m=0.5$ and the right one 
is for changing values of $m$ by keeping $c_2=1.5$ for all. Another model 
parameter $n$ is taken as $0.5$ for both cases. The figure demonstrates 
the increase in the deflection angle as the impact parameter decreases, then 
diverging to reach the observer at $\zeta=\zeta_p$. Additionally, according to 
the left plot as $c_2$ increases, the point of divergence shifts to higher 
values of impact parameter i.e., the light diverges at greater $\zeta_p$ value 
for the larger $c_2$, thereby approaching the characteristics of 
Schwarzschild BH. In contrast, it is evident from the right plot that for 
smaller values of $m$, the deflection angle remains close to that of the 
Schwarzschild case and shows a gradual deviation from the angle as $m$ 
increases. The deflection angle is seen to move towards the negative end 
typically as mentioned in the weak lensing case for larger $m$ and the smaller 
$c_2$ values respectively. Notably, Fig.~\ref{fig3} demonstrates that the 
Hu-Sawicki BHs possess smaller $\zeta_p$ values than that of the Schwarzschild 
BH for any value on model parameters. This suggests the induction of a 
weaker gravitational field of the Hu-Sawicki BHs due to modification as 
compared to the Schwarzschild BH.

\begin{figure}[!h]                          
	\centerline{
	\includegraphics[scale = 0.76]{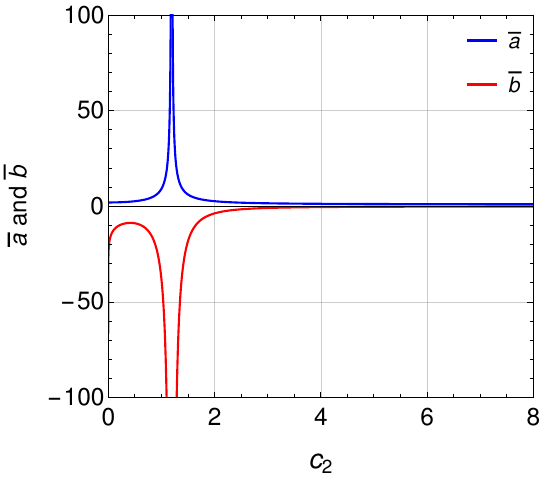}\hspace{0.7cm}
	\includegraphics[scale = 0.76]{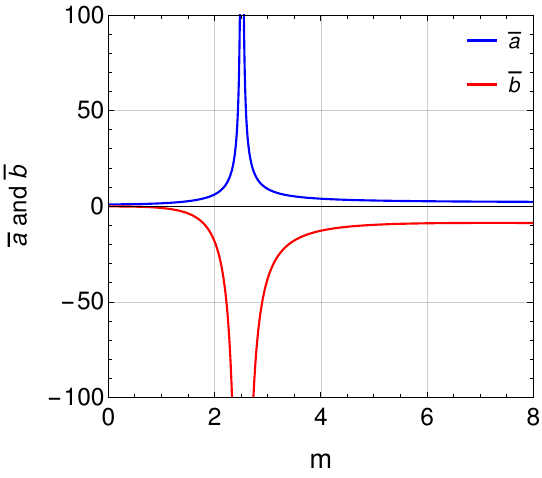}}\vspace{-0.2cm}
	\caption{Variation of $\bar{a}$ and $\bar{b}$ with model parameters 
$c_2$ and $m$ for $M=1$, $n=0.5$. The blue and red curves represent $\bar{a}$ 
and $\bar{b}$ respectively.}
	\label{fig4}
\end{figure}

The coefficients $\bar{a}$ and $\bar{b}$ rely on metric functions calculated 
at $r_p$ which in Schwarzschild's case are found to be equal to $1$ and 
$-0.400841$ respectively. To illustrate the behavior of coefficients $\bar{a}$ 
and $\bar{b}$, we present their graphical representation as functions of the 
model parameters $c_2$ and $m$ collectively in Fig.~\ref{fig4}. The figure 
illustrates that initially $\bar{a}$ increases with increasing the model 
parameters, reaches a peak, then falls and finally remains constant beyond 
certain parameter values. In contrast, the initial behavior of $\bar{b}$ is
different with respect to parameters $c_2$ and $m$. $\bar{b}$ increases 
initially with respect to $c_2$, but decreases initially for the parameter $m$. 
However, as in the case of the parameter $m$, $\bar{b}$ decreases after
a particular value of $c_2$ and then for both cases it increases after 
attaining a sufficiently smaller value. It becomes constant as the model 
parameters grow. This type of behavior of $\bar{a}$ and $\bar{b}$ 
is also found in the Simpson-Visser black-bounce spacetime as described in 
Ref.~\cite{2021_Tsuk}. 
                                     
\subsection{Lensing observables}
From the deflection angle calculated in the strong field limit, the properties 
of relativistic images can be obtained using a lens equation that relates the 
position of the source and images formed by the lensing object. The formation 
of the relativistic images is due to the spiraled light rays around the BH in 
the strong gravitational field and are significantly demagnified relative to 
the standard weak field images unless there is a strong alignment of the 
source, lens, and observer~\cite{Virbhadra_2000, Whisker_2005, Bozza_2004}. In 
the case of a highly aligned configuration (see Fig.~\ref{fig6}), the lens 
equation can be expressed as follows~\cite{2002_Bozza,2024_islum,2011_Eiroa, Kuang_2022, Virbhadra_2000,2009_jing}:  
\begin{equation} 
	\beta = \vartheta - \frac{d_{ls}}{d_{os}}\, \Delta \alpha_k,
	\label{eq52}
\end{equation}
where $\beta$ and $\vartheta$ represent the angular separation between the 
lens and the source, and the lens and the image respectively. 
$\Delta \alpha_k = \hat{\alpha} - 2k\pi$, with $k$ as a positive integer 
represents a winding number of light rays in multiple loops around the BH. 
$d_{ls}$ and $d_{os}$ are distances of the lens (BH) from the source and the 
observer respectively. Using Eqs.~\eqref{eq47}, \eqref{eq52} and the relation 
$\zeta \approx \vartheta d_{ol}$, the angular position $\vartheta_k$ of 
$k^{th}$ relativistic image can be expressed 
as~\cite{2024_islum,2011_Eiroa,2023_Kumar,Zhao_2017} 
\begin{equation} 
	\vartheta_k = \vartheta_k^0 + \Delta \vartheta_k,
	\label{eq53}
\end{equation}
where $\vartheta_k^0$ represents the image position corresponding to  
$\hat{\alpha} = 2k\pi$ when a photon winds complete $2k\pi$ around the BH, 
and is found as~\cite{2002_Bozza,2024_islum,Kuang_2022} 
\begin{equation} 
	\vartheta_k^0 = \frac{\zeta_p}{d_{ol}}(1 + e_k),
	\label{eq54}
\end{equation}
and
\begin{align} 
	\Delta \vartheta_k & = \frac{(d_{ol} + d_{ls})\, \zeta_p\, e_k}{\bar{a}\,d_{ls} d_{ol}}\,(\beta - \vartheta_k^0),
	\label{eq55}\\[8pt]
	e_k & = \exp\left(\frac{\bar{b} - 2k\pi}{\bar{a}}\right).
	\label{eq56}
\end{align}
Thus Eq.\eqref{eq53} can be approximated as~\cite{2002_Bozza,2024_islum,2009_jing,Whisker_2005} 
\begin{equation} 
	\vartheta_k = \vartheta_k^0 +\frac{\zeta_p e_k(\beta - \vartheta_k^0)d_{os}}{\bar{a}\,d_{ls}d_{ol}}.
	\label{eq57}
\end{equation}	
\begin{figure}[!h]                                                                 \centerline{
		\includegraphics[scale = 0.5]{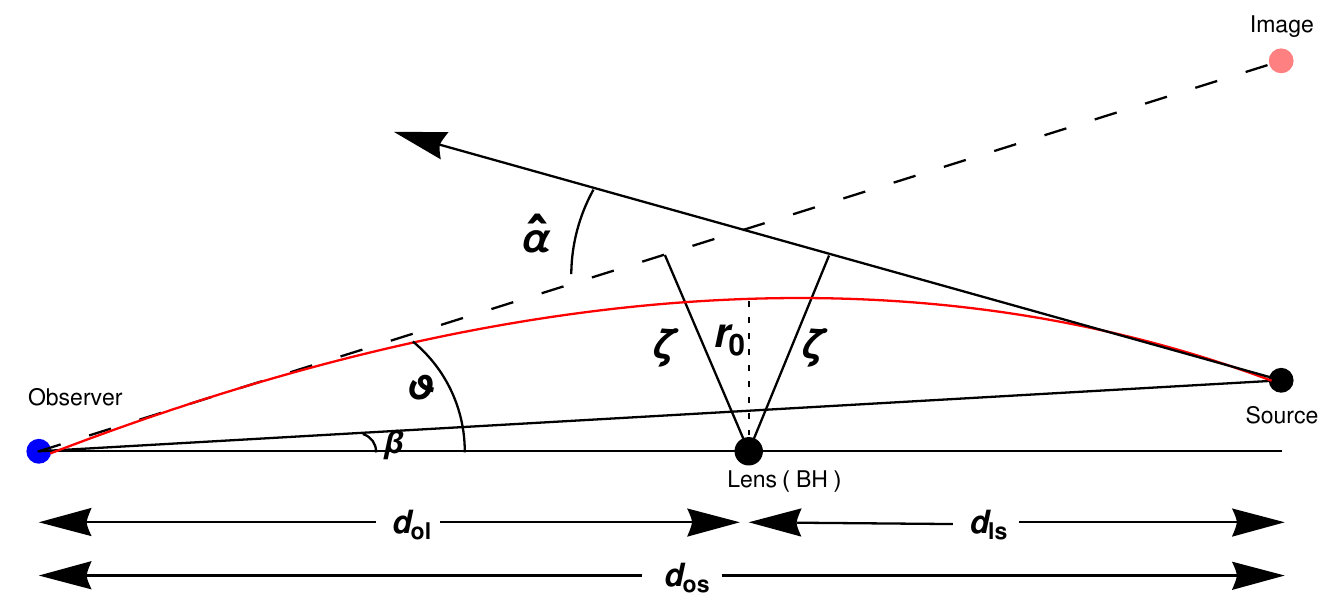}}\hspace{0.3cm}
	\caption{A schematic diagram of strong gravitational lensing.}
	\label{fig6}
\end{figure}
\noindent This equation links the position of relativistic images with the 
lensing coefficients $\bar{a}$, $\bar{b}$ and critical impact parameter 
$\zeta_p$. From Eq.~\eqref{eq54} we can evaluate $\vartheta_{\infty}$, the 
asymptotic position of a set of images. Since as $k \to \infty$, $e_k$ 
approaches zero, thereby the equation results in 
$\vartheta_{\infty}=\frac{\zeta_p}{d_{ol}}$~\cite{2024_islum,2009_jing}. 
Furthermore, treating $\vartheta_1$ as a single-loop outermost relativistic 
image and $\vartheta_{\infty}$ as a pack of all inner ones, the angular 
separation $s$ between these two can be obtained as
\cite{2002_Bozza,2020_islum,Ding_2011,2009_jing} 
\begin{equation} 
	s = \vartheta_1 - \vartheta_\infty = \vartheta_\infty \exp\left(\frac{\bar{b} - 2\pi}{\bar{a}}\right).
	\label{eq58}
\end{equation}
Another important characteristic observable in strong gravitational lensing is 
the magnification of relativistic images which is the inverse of the Jacobian 
determinant evaluated at the position of the image and is given for $k^{th}$ 
image as~\cite{2001_Bozza,2002_Bozza,2022_kumar, Whisker_2005} 
\begin{equation} 
	 \mu_k =\left. \left(\frac{\beta}{\vartheta}\,\frac{\partial \beta}{\partial \vartheta}\right)^{-1} \right|_{ \vartheta_k^0} \simeq \frac{e_k(1+e_k)d_{os}}{\bar{a}\, \beta\ d_{ls}\ d^2_{ol}}\, \zeta^2_p.
	\label{eq59}
\end{equation}
It is evident from the above equation that $\mu_k$ is an inverse function of 
$d^2_{ol}$. Therefore, it is quite small which makes the relativistic images 
usually dim except for $\beta\to 0$ because this limiting condition leads to 
an almost perfect alignment of the source, observer and lens. In conjunction 
with Eq.~\eqref{eq59} the ratio of magnification $r_{\text{mag}}$ of the 
outermost relativistic image at $\vartheta_1$ to remaining packed inner images 
at $\vartheta_\infty$ can be found as~\cite{Kuang_2022,2024_islum} 
\begin{equation} 
	r_{\text{mag}}=\frac{\mu_1}{\sum_{k=2}^{\infty}\mu_k} = \exp\left(\frac{2\pi}{\bar{a}}\right).
	\label{eq60}
\end{equation}
Interestingly, this observable is independent of mass and the distance between 
the BH and the observer. In the following subsection, we will consider 
observational data of two supermassive BHs $\text{Sgr A}^*$ and $\text{M87}^*$ 
to compute the values of lensing coefficients numerically predicted by the 
Hu-Sawicki BH. 

\subsection{Evaluation of observables from supermassive BHs}
We assume that the gravitational fields of the supermassive BHs 
$\text{Sgr A}^*$ at the centers of our galaxy and $\text{M87}^*$ in 
Messier 87 galaxy are characterized by the Hu-Sawicki BHs within the framework 
of $f(R)$ gravity theory. We compute the numerical values of key observables 
such as angular position $\vartheta_{\infty}$, angular separation $s$, and 
relative magnification $r_{\text{mag}}$ related to gravitational lensing in 
the strong field limit to investigate the impact of model parameters on the 
lensing effect. For this, we employ the mass $M = 4.3 \times 10^6  M_\odot$ of 
$\text{Sgr A}^*$ and its distance $d_{ol} = 8.35$ kpc from the Earth based on 
Refs.~\cite{Kuang_2022,Dong_2024,2008_Ghez,2019_Do,2009_Gill}.
\begin{table}[!h]
	\centering
	\caption{Estimation of numerical values of characteristic strong 
lensing observables along with lensing coefficients and corresponding impact 
parameter $\zeta_p$ for the BHs $\text{SgrA}^*$ and $\text{M87}^*$. Here $r_s$ 
refers to the Schwarzschild radius.}
	\vspace{0.2cm}
	\setlength{\tabcolsep}{10pt}
	\scalebox{0.9}{
		\begin{tabular}{c c c c c c c c c c c c c} \hline\\[-12pt]
			&~    & ~~~   $\text{SgrA}^*$~~~~~& ~~$\text{SgrA}^*$ &~~$\text{M87}^*$~~&~~$\text{M87}^*$ &~  ~&~~~~ &\\[3pt] 
			$m$~~ & ~~~ $c_2$~~~& $\vartheta_{\infty}(\mu \text{as})$& s $(\mu \text{as})$ &$\vartheta_{\infty}(\mu\text{as})$& s $(\mu \text{as})$ & ~~$r_{\text{mag}}$~~&~~$\zeta_p/r_s$~~&~~$\bar{a}$~~~~&~~~ $\bar{b}$\\[2pt] \hline \hline\\[-12pt]     		
     0  &  0               & 26.5972  & 0.033239   &  19.9473 & 0.024820  & 6.77155 & 2.59808 & 1.00000 & -0.40084  \\ [2pt]\hline\\[-12pt]
		& 0.8              & 21.7573  & 0.027190   &  16.3175 & 0.020392  & 4.08529 & 2.12530 & 1.65754 & -1.48220  \\ [5pt]  
	0.5	& 1.1              & 23.6806  & 0.029594   &  17.7600 & 0.022194  & 5.13907 & 2.31311 & 1.31766 & -0.55410  \\ [5pt] 
		& 1.5              & 24.9037  & 0.031123   &  18.6772 & 0.023341  & 5.81554 & 2.43265 & 1.04527 & -0.23514  \\ [5pt]       
		& 2.2              & 25.7683  & 0.032203   &  19.3256 & 0.024151  & 6.30025 & 2.50104 & 1.07481 & -0.27168  \\ [2pt]
		\hline\\[-12pt]  		    		  
		& 0.8              & 18.9546  & 0.023687   &  14.2155 & 0.017765  & 2.52765 & 1.85153 & 2.70229 & -5.75272  \\ [5pt]	 
	0.7	& 1.1              & 21.6264  & 0.027027   &  16.2193 & 0.020269  & 4.04836 & 2.11252 & 1.68721 & -1.24456  \\ [5pt]
		& 1.5              & 23.5491  & 0.029429   &  17.6613 & 0.022071  & 5.11085 & 2.30033 & 1.33646 & -0.59376  \\ [5pt]
		& 2.2              & 25.0413  & 0.031294   &  18.7804 & 0.023470  & 5.89223 & 2.44609 & 1.14923 & -0.29789  \\ [2pt]
		\hline\\[-12pt]
		& 0.8              & 16.4892  & 0.020607   &  12.3665 & 0.015454  & 0.91525 & 1.61070 & 7.46295 & -33.8099  \\ [5pt]
	0.9	& 1.1              & 19.5693  & 0.024456   &  14.6765 & 0.018341  & 2.88859 & 1.91157 & 2.36462 & -4.22514  \\ [5pt]
		& 1.5              & 22.0451  & 0.027550   &  16.5333 & 0.020662  & 4.27993 & 2.15341 & 1.59592 & -1.28673  \\ [5pt]
		& 2.2              & 24.1612  & 0.030195   &  18.1204 & 0.022645  & 5.45080 & 2.36012 & 1.25311 & -0.43259  \\[2pt]\hline \hline	
	\end{tabular}}
	\label{table1}
\end{table}
\begin{figure}[!h] 
	   \centerline{
		\includegraphics[scale = 0.61]{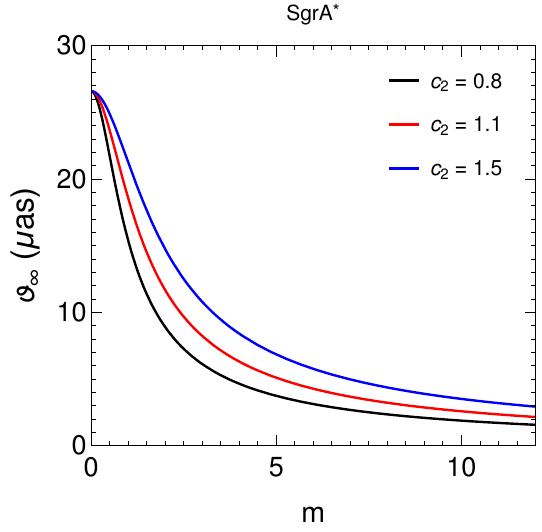}\hspace{0.3cm}
		\includegraphics[scale = 0.61]{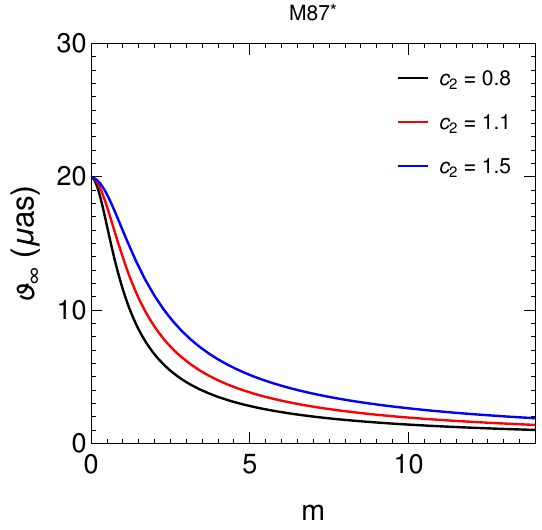}\hspace{0.3cm}
		\includegraphics[scale = 0.60]{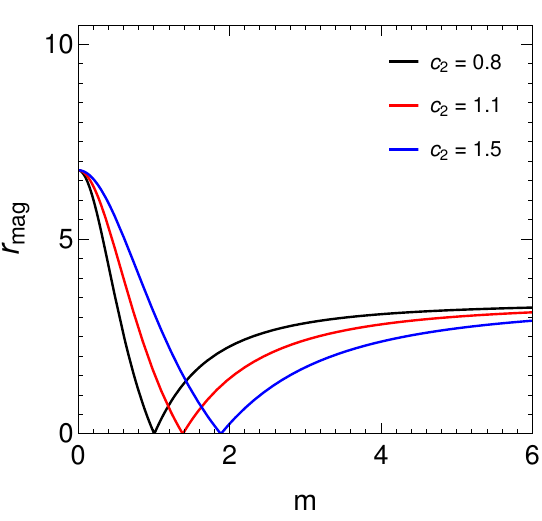}}\vspace{0.3cm}
         \centerline{     
		\includegraphics[scale = 0.61]{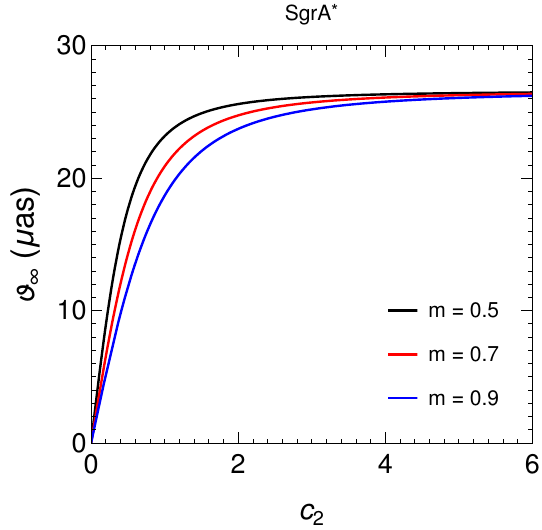}\hspace{0.3cm}
	        \includegraphics[scale = 0.61]{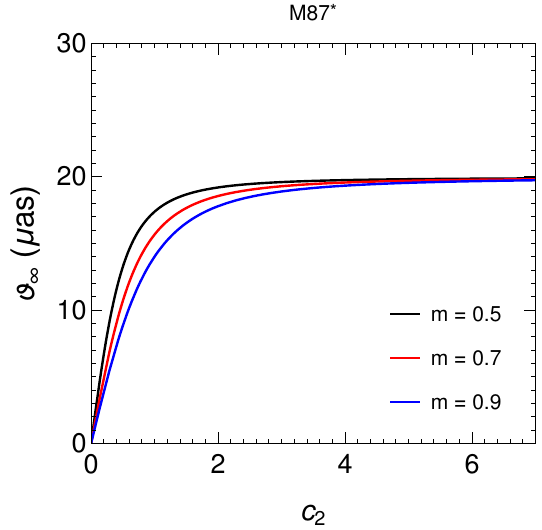}\hspace{0.3cm}
		\includegraphics[scale = 0.61]{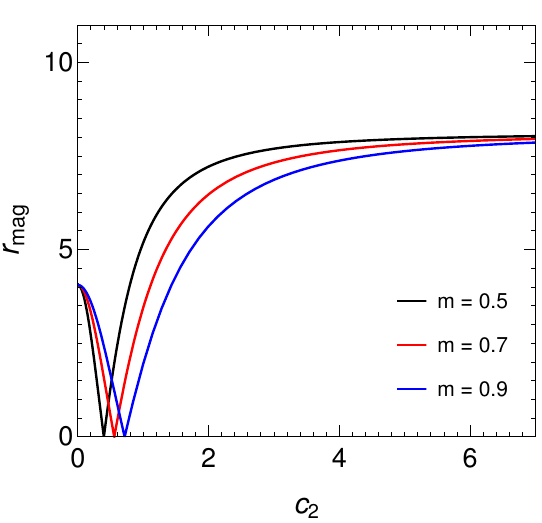}}\vspace{-0.3cm}
	\caption{Variation of strong lensing observables $\vartheta_{\infty}$ 
and $r_{\text{mag}}$ with respect to parameters $m$ and $c_2$ for BHs 
$\text{Sgr A}^*$ and $\text{M87}^*$. The left two plots present the change of 
$\vartheta_{\infty}$ for $\text{Sgr A}^*$, middle two plots depict the 
corresponding change for $\text{M87}^*$ while the right plots show the 
variation of relative magnitude $r_{\text{mag}}$. It is to be noted that 
$r_{\text{mag}}$ is not any BH specific.}
	\label{fig5}
\end{figure}
\begin{figure}[!h]                                                                 
	\centerline{
	\includegraphics[scale = 0.62]{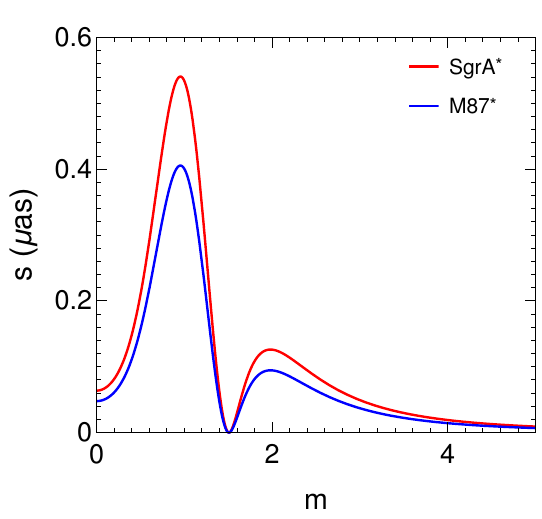}\hspace{0.7cm}
	\includegraphics[scale = 0.62]{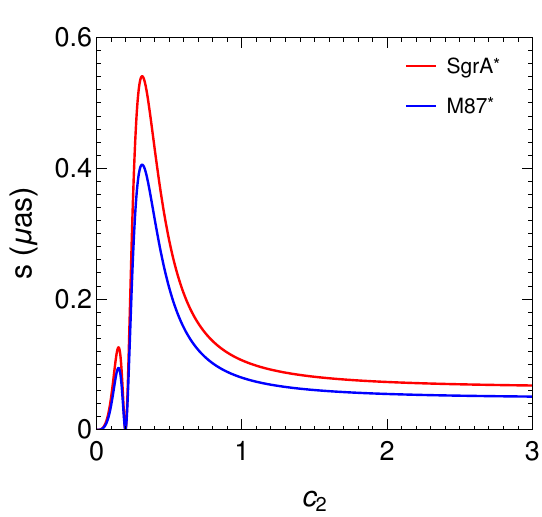}\vspace{-0.2cm}}
	\caption{Behavior of $s$ with parameters $m$ and $c_2$ for BHs 
$\text{Sgr A}^*$ and $\text{M87}^*$. The left panel illustrates the variation 
of $s$ with $m$ keeping $c_2$ fixed at 1.5, and the right panel shows the 
variation of $s$ with $c_2$ keeping $m = 0.5$.}
	\label{fig6}
\end{figure}
Similarly, in the case of $M87^*$, a mass $M = 6.5 \times 10^9 M_\odot$ and a 
distance $d_{ol} = 16.8$ Mpc are adopted following the data in the 
Refs.~\cite{Kuang_2022,AkiyamaSgrA_2022,Dong_2024}. The numerically calculated 
the characteristic lensing observables $\vartheta_{\infty}$, $s$ and 
$r_{\text{mag}}$ across different values of model parameters $m$ and $c_2$ for 
both the BHs are presented in Fig.~\ref{fig5} and Fig.~\ref{fig6}, and 
tabulate in Table~\ref{table1}. Fig.~\ref{fig5} shows that 
$\vartheta_{\infty}$ decreases with increasing $m$ and almost remains 
unaltered for larger $m$. Contrary to this, it increases steeply with 
increasing $c_2$ initially and becomes constant once $c_2$ attains a certain 
value. Meanwhile, the relative magnitude $r_{\text{mag}}$ displays almost a 
similar pattern in both cases of varying $m$ and $c_2$. It decreases first for 
a small range of the values of parameters and then rises to a fixed value 
across the greater range of parameters in both cases. It is seen that the fall 
and rise of $r_{\text{mag}}$ with parameter $m$ is shallower than that for the 
variation with $c_2$. In addition, Table~\ref{table1} highlights the 
differences in the values of $\vartheta_{\infty}$, $s$ and $r_{\text{mag}}$ 
produced by the Hu-Sawicki model in comparison to the Schwarzschild case. The 
table shows that the minimum angular position $\vartheta_{\infty}$ reaches 
16.4892 µas for $\text{Sgr A}^*$ and 12.3665 µas for $M87^*$ in the Hu-Sawicki 
case whereas $\vartheta_{\infty}$ in Schwarzschild case for these respective 
BHs are 26.5972 and 19.9473 respectively. Thus, one may find deviations 
$\Delta \vartheta_{\infty}$ as 9.7480 µas and 7.5808 µas for these two 
respective BHs. On the other hand, in Fig.~\ref{fig6} we jointly depict the 
behavior of observable $s$ with parameters $m$ and $c_2$ respectively for the
BHs $\text{SgrA}^*$ and $\text{M87}^*$. Observation reveals that $s$ shows a 
bimodal behavior initially following an attenuated pattern and then decreases 
as $m$ increases. Contrarily, in the case of $c_2$ variations, a bimodal 
pattern with increasing peak value is observed with a flattened behavior in 
the large parameter value. The figure also clearly indicates a small angular 
separation between $\text{SgrA}^*$ and $\text{M87}^*$ for both $m$ and $c_2$. 

\section{Conclusion}
\label{sec.5}
Gravitational lensing is a significant tool for probing the nature of BHs 
across various gravity theories. In this work, we have analyzed the 
gravitational lensing phenomenon in both the weak and strong field lensing 
regimes in the light of BHs governed by the Hu-Sawicki model in the framework 
of the $f(R)$ gravity theory. The deflection angle and the lensing properties 
of such BHs have been investigated in the weak field limit by employing the 
extended form of the GBT developed by Ishihara {\it et al.} 
\cite{Ishihara_2016} and in the strong field limit by employing a widely 
recognized technique proposed by V.~Bozza~\cite{2002_Bozza}. 

In the first part of our work, we investigate the bending angle of light in 
the weak field limit. The approach employed in this scenario is the extension 
of the GBT developed by Ishihara {\it et al.} to deduce the deflection angle 
of light in the weak field of the BH under study from the viewpoint of the 
receiver. This approach is independent of the asymptotic flatness of the 
spacetime in which it is employed. As a result, we applied this method to the 
asymptotically non-flat BH in the Hu-Sawicki model $f(R)$ gravity.
Thus, to examine the bending of light in this type of BHs we focus on 
how the deflection angle varies with the impact parameter along with the
variation of model parameters $c_2$ and $m$ through graphical analyses. It is 
seen from the analyses that there is a distinct deviation from the 
Schwarzschild case, particularly at larger impact parameters where the 
deflection angle becomes negative. Higher values of $c_2$ lead the deflection 
angle closer to the Schwarzschild behavior, though it remains negative at 
large impact parameters. Oppositely, lower values of $m$ allow the recovery 
of the Schwarzschild behavior and show near alignment with $m=0.05$ even at 
large impact parameter values. Our study has revealed that the model 
parameters have a significant effect on the deflection angle and give us an 
idea about the gravitational nature of the BHs. 

In the next part, we have analyzed how the deflection of a light ray around 
the Hu-Sawicki BHs depends on the impact parameter $\zeta$ by plotting 
$(\hat{\alpha} - \zeta)$ graph in the strong field limit by varying parameters 
$c_2$ and $m$ at $M=1, n=0.5$, and found that 
corresponding to a particular impact parameter the deflection angle increases 
with the decrease of the impact parameter value and shows divergence at the 
photon sphere as expected. Additionally, it is observed that deflection angles 
of Hu-Sawiki BHs move away from the Schwarzschild case as parameter $m$ 
increases but approach it with increasing the parameter $c_2$ taking on 
negative values at larger impact parameters similar to the weak lensing case. 
Of course, the negative values in the strong lensing limit naturally appear at 
smaller impact parameters compared to the weak lensing scenario. The occurrence 
of the negative deflection angle at large impact parameters suggests photon 
repulsion by the black hole's gravitational field, indicating a significant 
influence of the Hu-Sawicki $f(R)$ gravity on the gravitational lensing 
phenomenon. We have analyzed the behavior of strong lensing 
coefficients $\bar{a}$ and $\bar{b}$ in Hu-Sawicki spacetime and seen that 
$\bar{a}$ and $\bar{b}$ exhibit intricate behavior with stabilization at 
larger values of model parameters. \\
\indent We then extend our analysis to the supermassive BHs 
$\text{Sgr A}^*$ and M$87^*$ as the models of Hu-Sawicki BHs to numerical 
estimation of the basic observables $\vartheta_\infty$, $s$ and $r_\text{mag}$ 
related to the strong gravitational lensing effect. For this purpose, the mass 
and distance parameters of these objects as reported by EHT group
\cite{2009_Gill, AkiyamaM87_2019} are used. The results thus obtained are 
listed in Table~\ref{table1} as mentioned which tells us that the modification 
of BHs spacetime by Hu-Sawicki $f(R)$ gravity reduces the value of 
$\vartheta_{\infty}$ relative to GR (Schwarzschild case). The amount of 
reduction depends sensitively on the parameters of the model. Thus, our 
findings reveal that the lensing observables exhibit notable sensitivity to 
the variation in model parameters. Specifically, $\vartheta_\infty$ decreases 
as $m$ increases, while it experiences a steep rise with $c_2$ before 
stabilization. Contrarily, $s$ and $r_\text{mag}$ display intricate variation 
with gradual flattening across higher parameter values. Finally, we may note 
that our work provides significant evidence that the Hu-Sawicki model 
introduces notable gravitational lensing signatures and demonstrates 
considerable deviations from the Schwarzschild case. Indeed, our analysis 
highlights how gravitational lensing in both weak and strong regimes can be 
utilized to investigate BH metrics within modified gravity theories. 
These studies can be extended to wormhole backgrounds in different gravity 
theories as a future prospect.                                                                                                      

\section*{Acknowledgements}
UDG is thankful to the Inter-University Centre for Astronomy and Astrophysics
(IUCAA), Pune, India for the Visiting Associateship of the institute.

\bibliographystyle{apsrev}

\end{document}